\documentclass[prd,preprintnumbers,superscriptaddress,nofootinbib,floatfix,twocolumn,notitlepage]{revtex4-1}

\usepackage{textcase}
\usepackage{graphicx,epsfig,psfrag,amssymb,hyperref}
\usepackage{multirow}
\usepackage{color,graphicx,epsfig,psfrag,amsmath,empheq}
\usepackage{bm}
\usepackage{mathrsfs,amsfonts,, color}
\usepackage[caption=false]{subfig}
\usepackage{hepunits}

\newcommand{\Ref}[1]{Ref.\:\cite{#1}}
\newcommand{\Refs}[1]{Refs.\:\cite{#1}}
\newcommand{\Fig}[1]{Fig.\:\ref{#1}}

\newcommand{\Tab}[1]{Table~\ref{#1}}

\newcommand{\Sec}[1]{Sec.\:\ref{#1}}

\newcommand{\App}[1]{App.\:\ref{#1}}

\newcommand{\Eq}[1]{Eq.\:(\ref{#1})}

\newcommand{\be}{\begin{eqnarray}}
\newcommand{\ee}{\end{eqnarray}}
\def\lsim{\mathrel{\rlap{\lower4pt\hbox{\hskip 0.5 pt$\sim$}}
\raise1pt\hbox{$<$}}}

\newcommand{\tev}{\ensuremath{\mathrm{\: Te\kern -0.1em V}}\xspace}
\newcommand{\gev}{\ensuremath{\mathrm{\: Ge\kern -0.1em V}}\xspace}
\newcommand{\mev}{\ensuremath{\mathrm{\: Me\kern -0.1em V}}\xspace}

\def\invfb   {\ensuremath{\mbox{\,fb}^{-1}}\xspace}
\def\pythia     {\mbox{\textsc{Pythia}}\xspace}

\def\geant     {\mbox{\textsc{Geant}}\xspace}
\def\Dbar    {{\kern 0.2em\overline{\kern -0.2em \mathrm{D}}{}}\xspace}

\def\A {{\ensuremath{A^{\prime}}}\xspace}
\def\mee {{\ensuremath{m_{e^+e^-}}}\xspace}
\def\epem {{\ensuremath{e^+e^-}}\xspace}
\def\smdecay {{\ensuremath{D^{*0}\!\to D^0\gamma}}\xspace}
\def\sigdecay {{\ensuremath{D^{*0}\!\to D^0\A}}\xspace}
\def\Adecay {{\ensuremath{\A \!\to e^+e^-}}\xspace}

\def\sigdecaypi {{\ensuremath{D^{*0}\!\to D^0\pi^0(\gamma \A)}}\xspace}
\def\jpsi     {{\ensuremath{{J\mskip -3mu/\mskip -2mu\psi\mskip 2mu}}}\xspace}
\def\kstarbar    {{\kern 0.2em\overline{\kern -0.2em K}{}^{*0}}\xspace}

\begin{document}

\title{Dark photons from charm mesons at LHCb}

\author{Philip Ilten}
\email{philten@cern.ch}
\affiliation{Laboratory for Nuclear Science, Massachusetts Institute of Technology, Cambridge, MA 02139, U.S.A.}

\author{Jesse Thaler}
\email{jthaler@mit.edu}
\affiliation{Center for Theoretical Physics, Massachusetts Institute of Technology, Cambridge, MA 02139, U.S.A.}

\author{Mike Williams}
\email{mwill@mit.edu}
\affiliation{Laboratory for Nuclear Science, Massachusetts Institute of Technology, Cambridge, MA 02139, U.S.A.}

\author{Wei Xue}
\email{weixue@mit.edu}
\affiliation{Center for Theoretical Physics, Massachusetts Institute of Technology, Cambridge, MA 02139, U.S.A.}

\begin{abstract}
We propose a search for dark photons $\A$ at the LHCb experiment using the charm meson decay $D^*(2007)^0 \!\to D^0 \A$.
At nominal luminosity, $D^{*0} \!\to D^0 \gamma$ decays will be produced at about 700\,kHz within the LHCb acceptance, yielding over 5 trillion such decays during Run~3 of the LHC.  
Replacing the photon with a kinetically-mixed dark photon, LHCb is then sensitive to dark photons that decay as $A^{\prime}\!\to e^+e^-$.
We pursue two search strategies in this paper.
The displaced strategy takes advantage of the large Lorentz boost of the dark photon  and the excellent vertex resolution of LHCb, yielding a nearly background-free search when the $\A$ decay vertex is significantly displaced from the proton-proton primary vertex.
The resonant strategy takes advantage of the large event rate for $D^{*0} \!\to D^0 \A$ and the excellent invariant mass resolution of LHCb, yielding a background-limited search that nevertheless covers a significant portion of the $\A$ parameter space.
Both search strategies rely on the planned upgrade to a triggerless-readout system at LHCb in Run~3, which will permit identification of low-momentum electron-positron pairs online during data taking.
For dark photon masses below about 100\,MeV, LHCb can explore nearly all of the dark photon parameter space between existing prompt-\A and beam-dump limits.
\end{abstract}

\preprint{MIT-CTP 4702}
\maketitle

\section{Introduction}
\label{sec:Introduction}

Rare decays of mesons are a powerful probe of physics beyond the standard model (SM).
Precise measurements of branching fractions and decay kinematics indirectly constrain extensions of the SM by bounding symmetry-violating or higher-dimensional operators.
More directly, non-SM particles could be produced in meson decays when kinematically allowed, and depending on their lifetimes, these particles could yield striking signals with displaced vertices.
A well-motivated hypothetical particle is the dark photon \A which inherits a small coupling to the SM via kinetic mixing with the ordinary photon $\gamma$~\cite{Okun:1982xi,Galison:1983pa,Holdom:1985ag,Pospelov:2007mp,ArkaniHamed:2008qn,Bjorken:2009mm}.  
Indeed, some of the most stringent constraints on the properties of dark photons come from rare decays of mesons, 
including $\pi^0\!\to\gamma\A$~\cite{Bernardi:1985ny,MeijerDrees:1992kd,Astier:2001ck,Gninenko:2011uv,Adlarson:2013eza,Agakishiev:2013fwl,Adare:2014mgk,Batley:2015lha}, 
$\eta/\eta^{\prime}\!\to \gamma\A$~\cite{Bergsma:1985is,Gninenko:2012eq}, 
and $\phi \!\to\eta \A$~\cite{Archilli:2011zc,Babusci:2012cr}. 

\begin{figure}[t]
\includegraphics[width=\columnwidth]{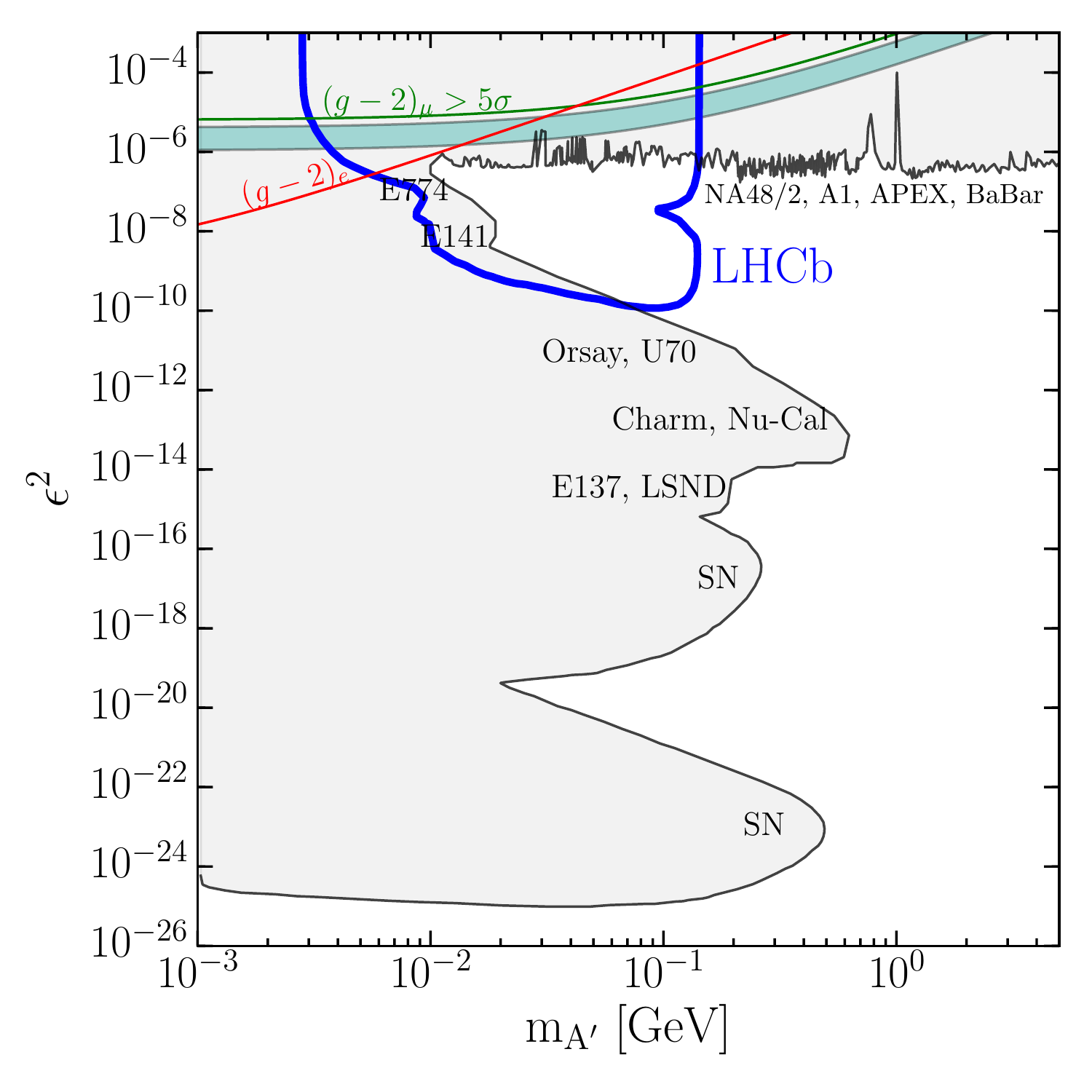}
\caption{Current bounds on dark photons with visible decays to SM states, adapted and updated from \Ref{Essig:2013lka}.  The upper bounds are from prompt-$A'$ searches while the wedge-shaped bounds are from beam-dump searches and supernova considerations.  The LHCb search region in \Fig{fig:lhcbbounds} covers most of the gap between these bounds for $m_\A \lesssim100\mev$, with a reach extending to $m_\A  \lesssim 140\mev$.  Anticipated limits from other planned experiments are shown in \Fig{fig:lhcbboundswithoverlay}.}
\label{fig:currentbounds}
\end{figure}

The minimal dark photon scenario involves a single broken $U(1)$ gauge symmetry, along with mixing between the \A and  SM hypercharge fields via the operator $F_{\mu\nu}^{\prime} B^{\mu \nu}$.
After electroweak symmetry breaking and diagonalizing the gauge boson kinetic terms, the dark photon gains a suppressed coupling to the ordinary electromagnetic current $J_{\rm EM}^\mu$, where the relevant terms in the Lagrangian are 
\be
\mathcal{L} \supset -\frac{1}{4}F'_{\mu\nu} F'^{\mu\nu} + \frac{1}{2}m_{A'}^2 A'_\mu A'^\mu + \epsilon e A_\mu' J_{\rm EM}^\mu.
\ee
This minimal scenario has two free parameters:  the dark photon mass $m_{\A}$ and the kinetic-mixing parameter $\epsilon$ (often reported in terms of $\epsilon^2$).
The constraints placed on dark photons in the $m_{A'}$--$\epsilon^2$ plane are shown in \Fig{fig:currentbounds} for $2m_e < m_{\A} < 5\gev$, assuming that the \A dominantly decays into visible SM states (see \Ref{Essig:2013lka} for a review).\footnote{There are also interesting searches where the dark photon decays invisibly to dark matter  \cite{Aubert:2008as,Batell:2009di,deNiverville:2011it,Wojtsekhowski:2012zq,Kahn:2012br,Izaguirre:2013uxa,Batell:2014mga,Kahn:2014sra,Izaguirre:2014bca,Izaguirre:2015yja}.}
For $\epsilon^2 \gtrsim 10^{-6}$, the most stringent bounds come from searches for prompt \A decays at collider and fixed-target experiments \cite{Batley:2015lha,Merkel:2011ze,Merkel:2014avp,Abrahamyan:2011gv}.
As $\epsilon$ decreases, the \A lifetime increases, while as $m_{\A}$ decreases, the lifetime and Lorentz boost factor both increase.
Therefore, the constraints obtained from beam-dump experiments exclude wedge-shaped regions in the  $m_{A'}$--$\epsilon^2$ plane~\cite{Bergsma:1985is,Konaka:1986cb,Riordan:1987aw,Bjorken:1988as,Bross:1989mp,Davier:1989wz,Athanassopoulos:1997er,Adler:2004hp,Bjorken:2009mm,Artamonov:2009sz,Essig:2010gu,Blumlein:2011mv,Gninenko:2012eq, Blumlein:2013cua}.
Also shown in \Fig{fig:currentbounds} are electron $g-2$ bounds \cite{Hanneke:2008tm,Giudice:2012ms,Endo:2012hp}\footnote{Since we follow the analysis in \Ref{Essig:2013lka}, we obtain more conservative bounds from $(g-2)_e$ than shown in \Ref{Endo:2012hp}.}, the preferred region to explain the muon $g-2$ anomaly \cite{Pospelov:2008zw}, and supernova bounds from cooling \cite{Dent:2012mx} and emissions \cite{Kazanas:2014mca}.
Anticipated limits from other planned experiments are shown later in \Fig{fig:lhcbboundswithoverlay}.

\begin{figure}[t]
\includegraphics[width=\columnwidth]{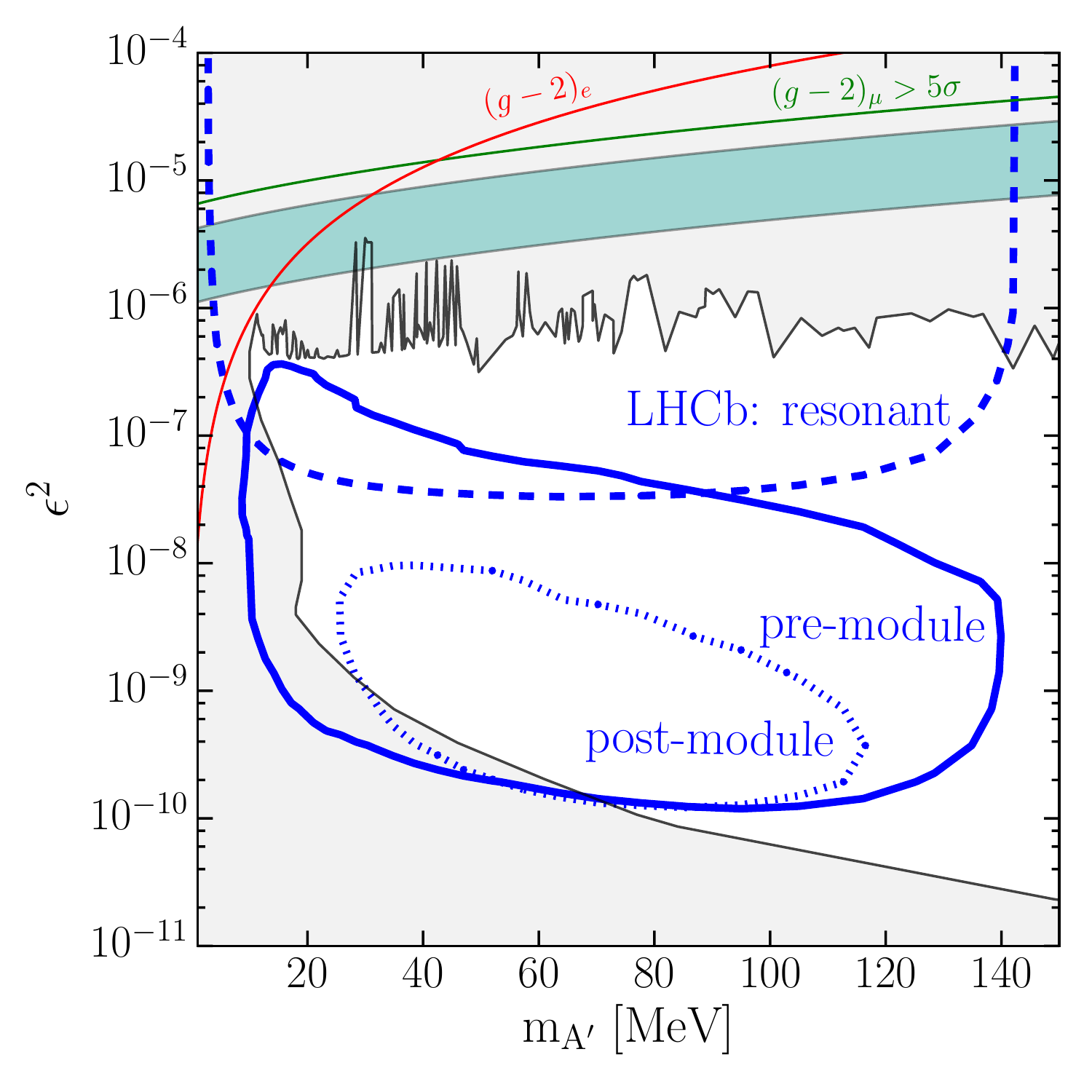}
\caption{Potential bounds from LHCb after Run 3, for both the displaced (pre-module, solid blue) and resonant (dashed blue) searches.  Also shown is an alternative displaced search strategy (post-module, dotted blue) that looks for $\A$ vertices downstream of the first tracking module.  
}
\label{fig:lhcbbounds}
\end{figure}

In this paper, we propose a search for dark photons through the rare charm meson decay
\be
\label{eq:maindecay}
\sigdecay, \quad \Adecay,
\ee
at the LHCb experiment during Run~3 of the LHC (scheduled for 2021--23).\footnote{Throughout this paper, $D^{*0} \equiv D^*(2007)^0$ and the inclusion of charge-conjugate processes is implied.}
The goal of this search is to explore the region between the prompt-\A and beam-dump limits for the range $m_{\A} \in [10,100] \mev$, which roughly includes $\epsilon^2 \in [10^{-10},10^{-6}]$.
Reaching such small values of $\epsilon^2$ is only possible for decays where the yield of the corresponding SM process (i.e.\ replacing $\A$ with $\gamma$) is at least $\mathcal{O}(10^{10})$.  Within the LHCb acceptance, over five trillion \smdecay decays will be produced in proton-proton ($pp$) collisions at 14\tev during Run~3, making this decay channel a suitable choice.  

The range of $m_{\A}$ values that is in principle accessible in this search is $m_{\A} \in [2m_e,\Delta m_D]$, where 
\cite{PDG2014}
\be
\Delta m_D \equiv m_{D^{*0}} - m_{D^0} = 142.12 \pm 0.07\mev.
\ee
The proximity of $\Delta m_D$ to $m_{\pi^0}$ leads to phase-space suppression of the decay $D^{*0}\!\to D^0\pi^0$, which results in a sizable branching fraction of about 38\% for the decay \smdecay.\footnote{This explains why we choose the decay \sigdecay instead of $D^*(2010)^+\!\to D^+ \A$, since the corresponding branching fraction $D^{*+}\!\to D^+ \gamma $ is only 1.6\%.}  The small value of $\Delta m_D$, however, also leads to typical electron momenta of $\mathcal{O}({\rm GeV})$ within the LHCb acceptance.  
Therefore, the planned upgrade to a triggerless-readout system employing real-time calibration at LHCb in Run~3~\cite{LHCb-TDR-016}---which will permit identification of relatively low-momentum $e^+e^-$ pairs online during data taking---will be crucial for carrying out this search.   

To cover the desired dark photon parameter space, we employ two different search strategies, shown in \Fig{fig:lhcbbounds}.  The \textit{displaced search}, relevant at smaller values of $\epsilon^2$, looks for an $\A\!\to e^+e^-$ decay vertex that is significantly displaced from the $pp$ collision.  This search benefits from the sizable Lorentz boost factor of the produced dark photons and the excellent vertex resolution of LHCb.  Our main displaced search looks for $\A$ decays within the beam vacuum upstream of the first tracking module (i.e.~pre-module), where the dominant background comes from misreconstructed prompt $D^{*0} \to D^0 e^+e^-$ events.\footnote{We thank Natalia Toro for extensive discussions regarding this background.}  Because the \A gains a transverse momentum kick from $pp$ collisions, the \A flight trajectory intersects the LHCb detector, making it possible to identify displaced $e^+ e^-$ pairs with smaller opening angles than the HPS experiment \cite{Moreno:2013mja}.  We also present an alternative displaced search for $\A$ decays downstream of the first tracking module (i.e.~post-module), where the dominant background comes from $D^{*0} \to D^0 \gamma$ events with $\gamma \to e^+e^-$ conversion within the LHCb material.  

The \textit{resonant search}, relevant at larger values of $\epsilon^2$, looks for an $\A\!\to e^+e^-$ resonance peak over the continuum SM background.  This search benefits from the large yield of \sigdecay decays during LHC Run 3, which is larger than the $A'$ yield in fixed-target experiments like MAMI/A1~\cite{Merkel:2011ze,Merkel:2014avp} and APEX~\cite{Abrahamyan:2011gv}.  Furthermore, the narrow width of the $D^{*0}$ meson, which is less than the detector invariant-mass resolution, provides kinematical constraints that can be used to improve the resolution on \mee.  This resonant search can also be employed for non-minimal dark photon scenarios where the \A might also decay invisibly into dark matter, shortening the $\A$ lifetime.  In that case, the anticipated limits in \Fig{fig:lhcbbounds} would roughly apply to the combination $\epsilon^2 \times \mathcal{B}(\Adecay)$.

The remainder of this paper is organized as follows. In \Sec{sec:dstar}, we estimate the \A signal and SM background cross sections, extracting the $D^{*0}$ production rate and $D^0$ decay modes from an event generator and estimating the $D^{*0}$ decay rates using a simple operator analysis.  In \Sec{sec:lhcb}, we describe the LHCb detector and charged-particle tracking, provide the selection requirements applied to $D^0$ and $D^{*0}$ meson candidates, and derive the \A mass resolution.  We present the pre-module displaced \A search in \Sec{sec:reach_dis}, a post-module variant in \Sec{sec:reach_dis_post}, and the resonant \A search in \Sec{sec:reach_res}.  Possible improvements are outlined in \Sec{sec:improve} and a comparison to other experiments (especially HPS) is given in \Sec{sec:other}.  We summarize in \Sec{sec:conclude} and discuss how the LHCb dark photon search strategy might be extended above the $\Delta m_D$ threshold.

\section{Signal and Background Rates}
\label{sec:dstar}

Dark photon production in $D^{*0}$ meson decays proceeds mainly via
\begin{align}
\sigdecay, \qquad \sigdecaypi, \label{eq:signalthroughpi}
\end{align}
though the low-energy photon in the latter decay is unlikely to be detected at LHCb.   Here and throughout, we use the notation $X(YZ)$ to mean $X\!\to YZ$ in a subsequent decay.
Because $\Delta m_D < 2m_{\mu}$, $A' \to e^+ e^-$ is the only relevant visible decay channel.

The dominant backgrounds to the pre-module displaced \A search (\Sec{sec:reach_dis}) are $D^{*0}\!\to D^0 e^+e^-$ and $D^{*0}\!\to D^0\pi^0(\gamma e^+e^-)$, where the $e^+e^-$ pair is misreconstructed as being displaced due to a hard electron scatter in material.  
These backgrounds can be highly suppressed by requiring that the $e^+e^-$ kinematics are consistent with a displaced $\A$ vertex occurring in the proper decay plane.
The dominant background to the post-module displaced $A'$ search (\Sec{sec:reach_dis_post}) is \smdecay, where the $\gamma$ converts into an $e^+e^-$ pair during interactions with the detector material.
This background can be highly suppressed by requiring that the $e^+e^-$ vertex position is not consistent with the location of any detector material.
The dominant backgrounds to the resonant search (\Sec{sec:reach_res}) are again $D^{*0}\!\to D^0 e^+e^-$ and $D^{*0}\!\to D^0\pi^0(\gamma e^+e^-)$, where the \A has been replaced by an off-shell $\gamma^*$.
The first background is irreducible, making the resolution on \mee the driving factor in the resonant search reach. 

\subsection{$D^{*0}$ Meson Production}

We simulate $D^{*0}$ production in $pp$ collisions at a center-of-mass energy of 14\tev using \pythia~8.201~\cite{Sjostrand:2014zea} with the default settings.  Since a large fraction of charm quarks are produced from gluon splitting and since we need to model forward physics at small transverse momentum $p_{\rm T}$, we run all soft QCD processes in \pythia\ (i.e.\ \texttt{SoftQCD:all = on}).  While the $D^{*0}$ production cross section is not yet known at 14\tev, the result obtained using \pythia for the inclusive $pp\!\to D^{*+}$ production cross section at 7\tev agrees with the measured value by LHCb\,\cite{LHCb-PAPER-2012-041} to within about 5\%.\footnote{\label{footnote:charmxsec}During the final preparation of this article, LHCb presented the first prompt charm cross section measurement at 13\tev~\cite{LHCb-PAPER-2015-041}.  Based on this result, we estimate that the relevant cross section for determining the dark photon reach should be about 20\% higher than the one used in this paper.}  Since \pythia does not record the spin of the $D^{*0}$ mesons, they are treated as unpolarized in this analysis.

To define the fiducial region, we require the $D^0$ meson to satisfy the following transverse momentum and pseudorapidity requirements:
\begin{equation}
p_{\rm T}(D^0) > 1 ~\GeV, \quad 2 < \eta(D^0) < 5.
\label{eq:cut0}
\end{equation}
Note that this requirement is placed on the $D^0$ meson, not on the $D^{*0}$, to suppress backgrounds to the $D^0$ component of the signal. 
The $D^{*0}$ production cross section within this fiducial region is
\be
\label{eq:baselinexsec}
\sigma(pp \to D^{*0}\!\to D^0_{\rm fid}) = 0.95~\text{mb},
\ee
excluding secondary production of $D^{*0}$ mesons from $b$-hadron decays.  It may be possible to make use of some secondary decays; in this analysis, however, we require that the \A originates from the $pp$ collision to suppress backgrounds (see \Sec{subsec:natalia}).  

The nominal instantaneous luminosity expected at LHCb during Run~3 is 2\:nb$^{-1}$ per second~\cite{LHCb-TDR-016}, which will produce $D^{*0}$ mesons at a rate of almost 2\:MHz (equivalently, \smdecay at 0.7\:MHz).  Assuming an integrated luminosity of 15\invfb in Run~3,\footnote{The length of Run~3 is scheduled to be about the same as Run~1.  LHCb collected a total of 3\,fb$^{-1}$ in Run~1.  The instantaneous luminosity will be five times higher in Run~3.  Therefore, assuming the LHC performance is the same (including the slow ramp up), this gives an estimate of 15\,fb$^{-1}$ in Run~3.} this results in an estimated yield of 14 trillion $D^{*0}$ mesons produced within this fiducial region, or 
\be
\label{eq:Nbench}
N ( D^{*0}  \rightarrow D^0 \gamma) = 5.4 \times 10^{12},
\ee
which we use as the baseline for our estimated reach.

\subsection{$D^{*0}$ Meson Decays}
\label{subsec:dstardecays}

The $D^{*0}$ meson is an $I(J^P) = \frac{1}{2}(1^-)$ state with a mass of $2006.96 \pm 0.10~\MeV$ and a width less than $2.1~\MeV$.  It decays promptly mainly into two final states with branching fractions of
\begin{align}
\mathcal{B}(D^{*0} \to D^0 \pi^0) &= (61.9 \pm 2.9)\%, \\
\mathcal{B}(D^{*0} \to D^0 \gamma) &= (38.1 \pm 2.9)\%,
\end{align}
where the $D^0$ meson is a $\frac{1}{2}(0^-)$ state~\cite{PDG2014}.  
As mentioned above, $D^{*0}\!\to D^0e^+e^-$ is the dominant background to the pre-module displaced search as well as to the resonant search.  To our knowledge, this branching fraction has not yet been measured; therefore, we will estimate the rate for this decay using an operator analysis.  This same approach is used to determine the \sigdecay rate.

To calculate these $D^{*0}\!\to D^0$ transition amplitudes, we must first determine the $\langle D^{*0} | J_{\rm EM}^\mu | D^0 \rangle$ matrix element.  By parity, time reversal, and Lorentz invariance, this transition dipole matrix element can be written in the form
\be
 \label{eq:DstarDmatrixelement}
\langle D^{*0} | J_{\rm EM}^\mu | D^0 \rangle = \mu_{\rm eff}(k^2) \, \epsilon^{\mu \alpha \beta \lambda} v_\alpha k_\beta \epsilon_\lambda,
\ee
where $v_\alpha$ is the four velocity of the $D^{*0}$ meson, $k_\beta$ is the momentum flowing out of the current, and $\epsilon_\lambda$ is the polarization of the $D^{*0}$ meson.  Here, $\mu_{\rm eff}$ is a $k$-dependent effective dipole moment, whose value could be determined using a simple quark model (see, e.g., \Ref{Miller:1988tz}) or using a more sophisticated treatment with heavy meson chiral perturbation theory (see, e.g., \Ref{Stewart:1998ke}).  For our purposes, we simply need to treat $\mu_{\rm eff}$ as being roughly constant over the range $k^2 \in [0, \Delta m_D^2]$, which is a reasonable approximation given that $\Delta m_D^2 < \Lambda_{\rm QCD}^2$.  (Indeed, this relation is always satisfied in the heavy charm quark limit, where $\Delta m_D \propto \Lambda_{\rm QCD}^2 / m_{c}$.)  The precise value of $\mu_{\rm eff}$ is irrelevant for our analysis since it cancels out when taking ratios of partial widths.

Using \Eq{eq:DstarDmatrixelement}, we estimate the decay rate for \smdecay within the SM and in the $\Delta m_D \ll m_D$ limit to be
\be
\Gamma(\smdecay) = \frac{\alpha_{\rm EM}}{3} \mu_{\rm eff}^2 \Delta m_D^3,
\ee
where $\alpha_{\rm EM} = e^2 / 4 \pi$.
To calculate the $D^{*0}\!\to D^0 e^+ e^-$ decay rate, the off-shell photon propagator must be included.  In the $m_e = 0$ limit, the amplitude for this process is
\be
\label{eq:ampDee}
\hspace{-0.15in}|\mathcal{M}_{D^{*0}\! \to D^0 e^+ e^-}|^2 = - \frac{2 e^4 \mu_{\rm eff}^2}{3} \! \left[ 1 \!- \!\frac{( k_1 \cdot v)^2\!\! + \!( k_2 \cdot v)^2}{k_1 \cdot k_2} \right]\!\!,\,\,\,
\ee
where $k_1$ and $k_2$ are the electron and positron momenta.
The ratio of partial widths is determined numerically to be
\be
\frac{\Gamma(D^{*0}\!\to D^0 e^+ e^-)}{\Gamma(D^{*0}\! \to D^0 \gamma)} = 6.4 \times 10^{-3}.
\ee 
Since the dark photon also couples to $J_{\rm EM}^\mu$, we use \Eq{eq:DstarDmatrixelement} to calculate the \sigdecay decay rate.  The ratio of partial widths is
\be
\label{eq:Dstar2DAprime}
\frac{\Gamma(\sigdecay)}{\Gamma(\smdecay)} =  \epsilon^2 \, \Bigl(1 - \frac{m_{\A}^2}{\Delta m_D^2} \Bigr)^{3/2},
\ee
where we assume $m_{\A}, \Delta m_D \ll m_D$.  
This expression has the expected kinetic-mixing and phase-space suppressions.  Since the $D^{*0}$ meson is treated as unpolarized in \pythia, we ignore spin correlations in the subsequent $\A\!\to e^+ e^-$ decay.\footnote{As a technical note, to generate \sigdecay events, we reweight a sample of \smdecay events from \pythia.  In particular, we implement \sigdecay in the $D^{*0}$ meson rest frame, boost to match the $D^{*0}$ kinematics from \pythia, and then boost the $D^0$ decay products to account for the altered $D^0$ momentum.  A similar strategy is employed for generating all other decays in our study.}

\subsection{Rare $\pi^0$ Decays}
\label{subsec:piondecay}

To determine the \sigdecaypi decay rate in \Eq{eq:signalthroughpi}, we start by estimating the rate of the decay $\pi^0\!\to\gamma \A$ using the SM effective Lagrangian
\be
\label{eq:piFF}
\mathcal{L} = \frac{\alpha_{\rm EM}}{2 \pi f_\pi} \pi^0 \epsilon^{\mu \nu \rho \sigma }  F_{\mu \nu} F_{\rho \sigma},
\ee
where $f_\pi$ is the pion decay constant and the pion form factor is ignored.  The dark photon is accounted for by making the replacement
\be
F_{\mu \nu} \to F_{\mu \nu} + \epsilon F'_{\mu \nu},
\ee
which leads to the ratio of partial widths
\begin{equation}
\frac{\Gamma(\pi^0\!\to \gamma \A)}{\Gamma(\pi^0 \to \gamma \gamma)} = 
   2  \epsilon^2 \left( \frac{ m_\pi^2 - m_{\A}^2} { m_\pi^2} \right)^3.
\end{equation}

The same effective Lagrangian can also be used for the SM decay $\pi^0 \!\to \gamma e^+ e^-$.  The amplitude is
\begin{eqnarray}
|\mathcal{M}_{\pi^0 \to \gamma e^+ e^-}|^2 &=&
\frac{4  \alpha_{\rm EM}^3}
{ \pi f_\pi^2  m_{\gamma e^-}^2}
 \bigg( m_{\pi^0}^4  + 2 m_{\gamma e^-}^4 + m_{\epem}^4
      \nonumber\\
&& \hspace{-0.8in} ~ + 2 m_{\gamma e^-}^2 m_{\epem}^2 - 2 m_{\pi^0}^2 (m_{\gamma e^-}^2+m_{\epem}^2)
       \bigg).
\end{eqnarray}
The ratio of partial widths is obtained numerically to be
\be
\frac{\Gamma(\pi^0 \!\to \gamma e^+ e^-)}{\Gamma(\pi^0\! \to \gamma \gamma)} =  0. 012,
\ee
which agrees with the nominal value for this ratio~\cite{PDG2014}.

\subsection{Dark Photon Decays}

\begin{figure}[t]
\includegraphics[width=0.96\columnwidth]{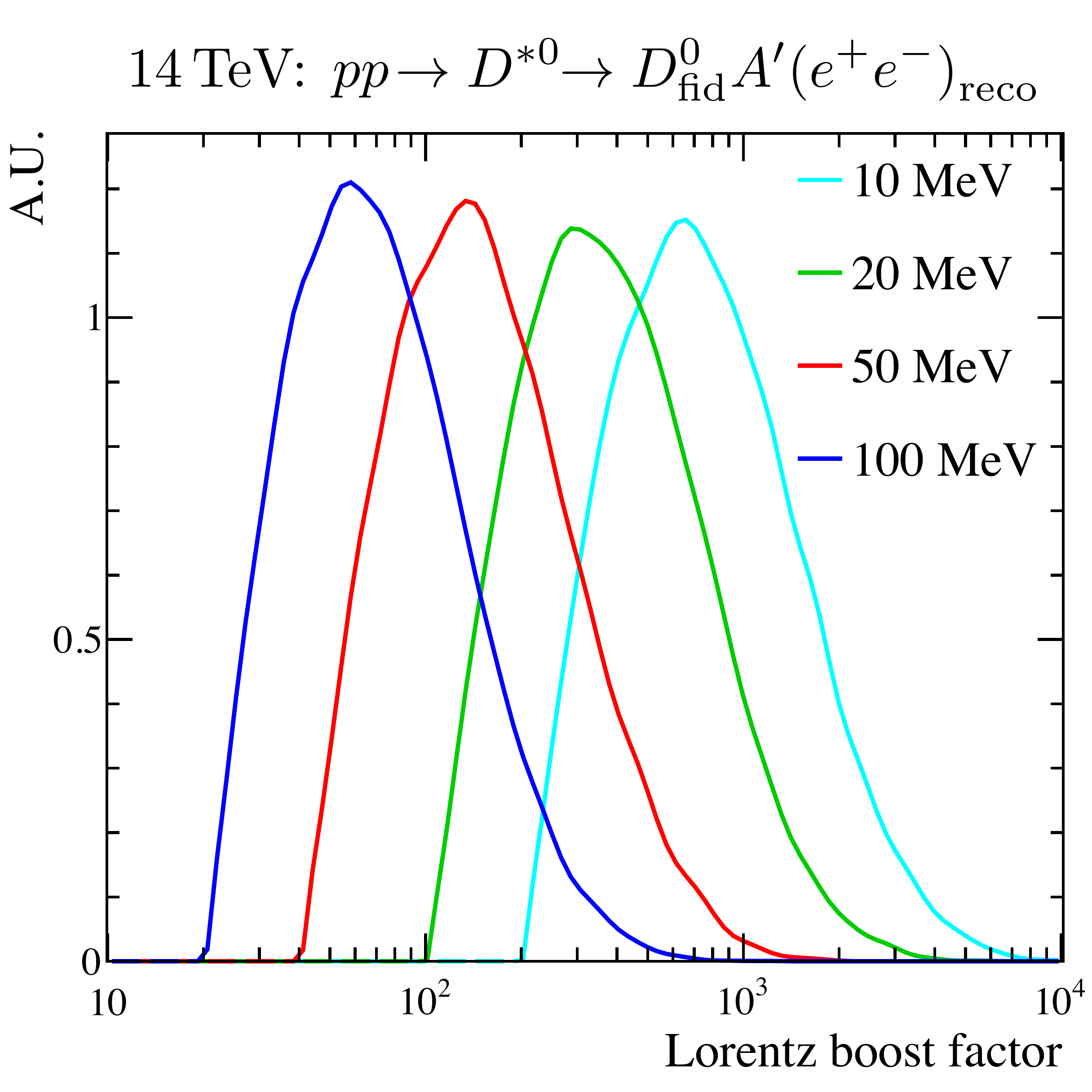}
\caption{Dark photon Lorentz boost factors for $m_\A = \{10,20,50,100\}~\MeV$.  These factors are independent of $\epsilon^2$. 
 }  
\label{fig:spc_gamma}
\end{figure}

\begin{figure*}[t]
\includegraphics[width=0.48\textwidth]{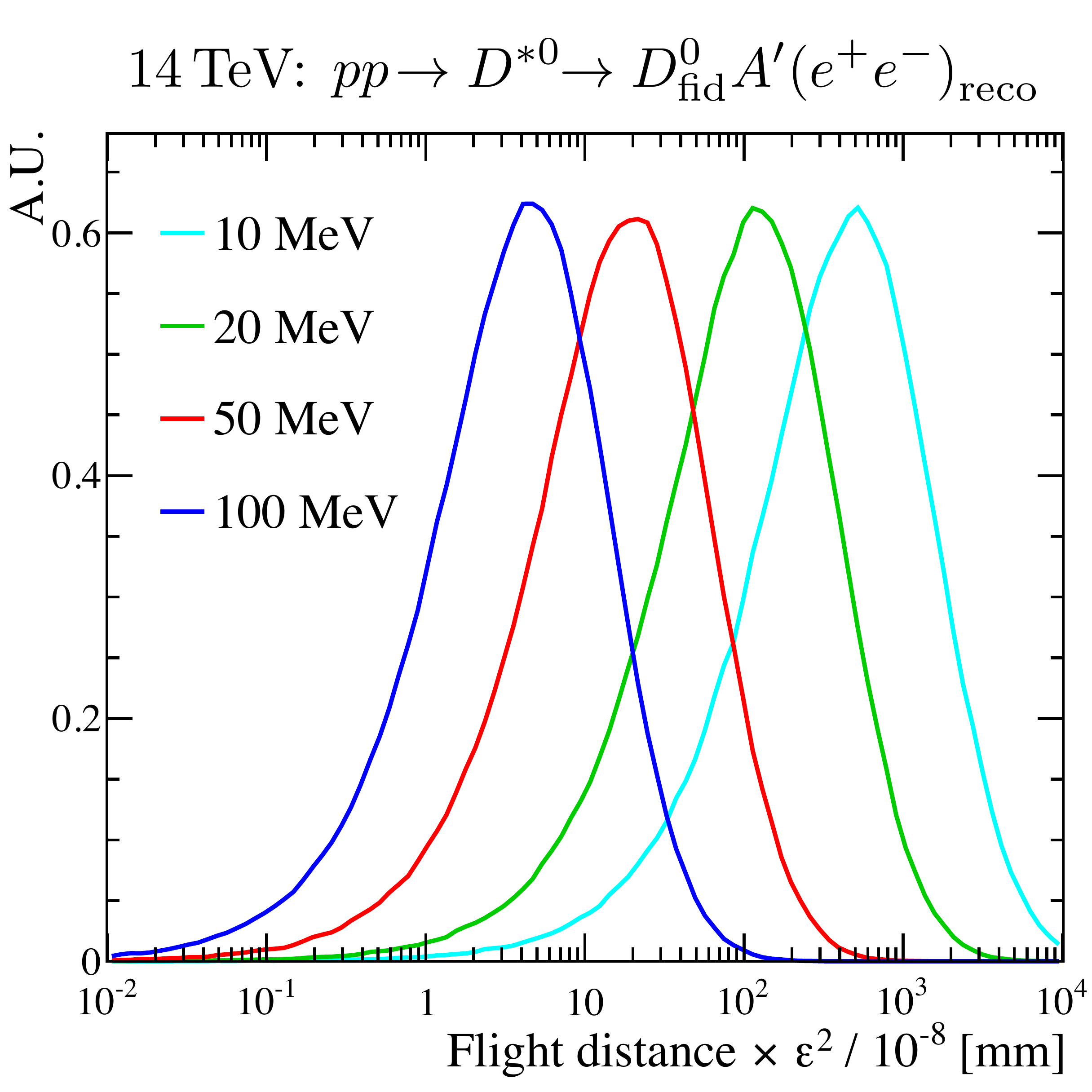}
$\quad$
\includegraphics[width=0.48\textwidth]{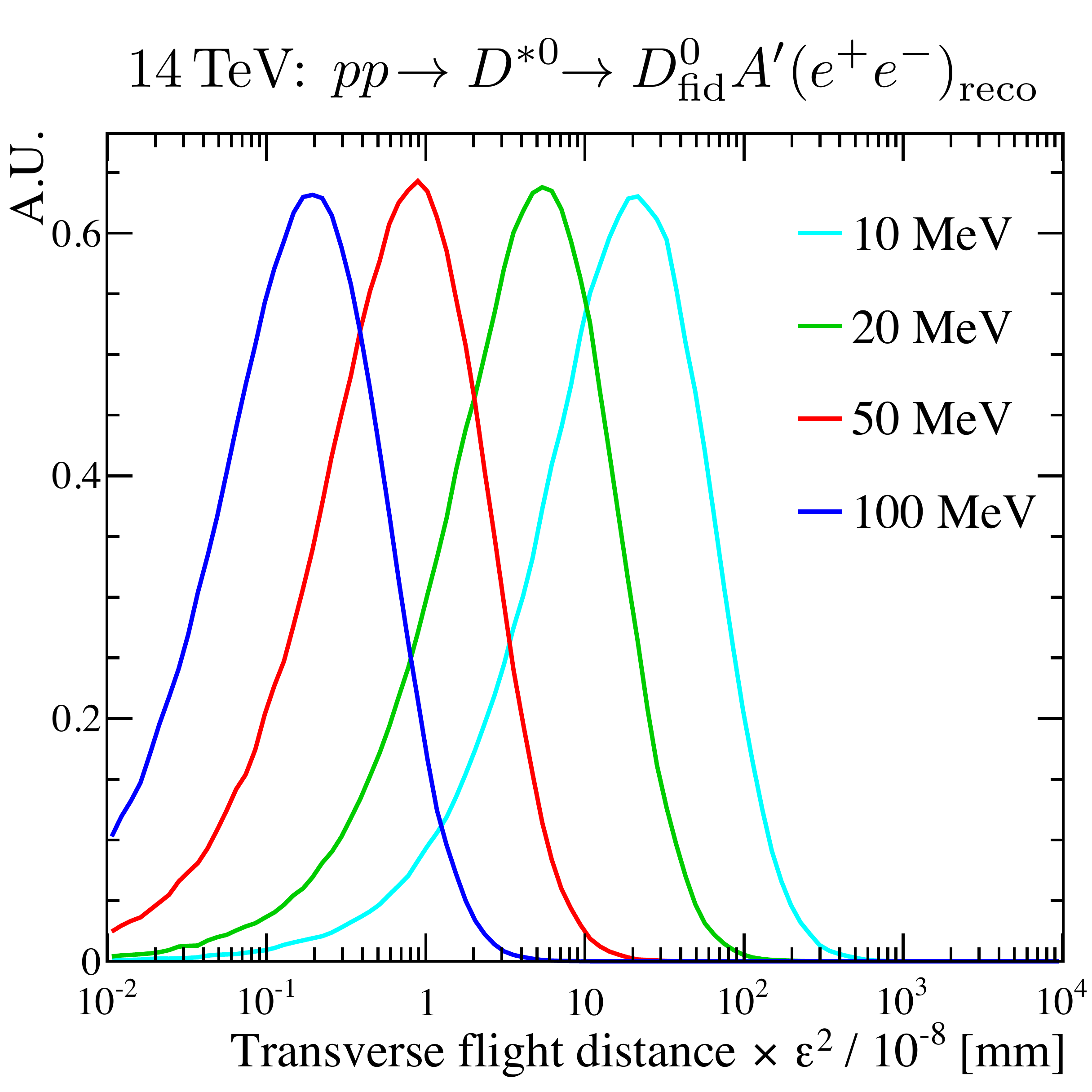}
\caption{Flight distance distributions for $m_{\A} = \{10,20,50,100\}$\mev showing (left) $\ell \times\left(\epsilon^2/10^{-8}\right)$ and (right) $\ell_{\rm T} \times \left(\epsilon^2/10^{-8}\right)$.
}  
\label{fig:spc_L}
\end{figure*}

Assuming the only allowed decay mode  is $\A\!\to e^+ e^-$, the total width of the \A is
\begin{equation}
\label{eq:gamma_a}
\Gamma_{\A} = \frac{\epsilon^2  \alpha_{\rm EM}} {3} m_{\A} 
      \left( 1 + 2 \frac{m_e^2}{ m_{\A}^2}  \right) 
      \sqrt{ 1 - 4 \frac{m_e^2}{ m_{\A}^2}  }. 
\end{equation}
In the lab frame, the mean flight distance of the dark photon is approximately
\begin{equation}
   \ell_{A'} \simeq   16\,\mathrm{mm} \, \left(\frac{\gamma_{\rm boost}}{10^2} \right) \left( \frac{ 10^{-8} }
      { \epsilon^2  }  \right)  \left( \frac{ 50~\mathrm{MeV} } 
      { m_{A'} } \right),
\end{equation}
where $\gamma_{\rm boost}$ is the Lorentz boost factor.  
In \Fig{fig:spc_gamma} we show some example spectra of \A boost factors from simulated \sigdecay decays, where both electrons are required to satisfy $|\vec{p}_e| \equiv p_e > 1\gev$ and $2 < \eta_e < 5$ so that they can be reconstructed by LHCb (see \Sec{sec:tracks} below).  
The \A inherits sizable momentum from the $D^{*0}$ meson, leading to $\gamma_{\rm boost}$ factors that reach $\mathcal{O}(10^4)$. 
The corresponding spectra for the total ($\ell$) and transverse ($\ell_{\rm T}$) flight distance of the \A are shown in \Fig{fig:spc_L}.  For $\epsilon^2 \lesssim 10^{-7}$ the displacement between the \A and $pp$-collision vertices is resolvable by LHCb.

\section{Baseline \texorpdfstring{LHC\MakeLowercase{b}}{LHCb} Selection}
\label{sec:lhcb}

The LHCb detector is a single-arm spectrometer covering the forward region of $2<\eta <5$~\cite{Alves:2008zz,Aaij:2014jba}.
The detector, which was built to study the decays of hadrons containing $b$ and $c$ quarks, includes a high-precision tracking system capable of measuring charged-particle momenta with a resolution of about 0.5\% in the region of interest for this search.\footnote{The precision of electron momentum measurements is limited by bremsstrahlung radiation; see \Sec{subsec:aprimereco}.}   
The silicon-strip vertex locator (VELO) that surrounds the $pp$ interaction region measures heavy-flavor hadron lifetimes with an uncertainty of about 50\,fs~\cite{LHCbVELOGroup:2014uea}.
Different types of particles are distinguished using information from two ring-imaging Cherenkov (RICH) detectors~\cite{LHCb-DP-2012-003}, a calorimeter system, and a system of muon chambers~\cite{LHCb-DP-2012-002}. 
Both the momentum resolution and reconstruction efficiency are $\mathcal{O}(10)$ times worse for neutral particles than for charged ones. For this reason, the analysis strategy outlined below is based entirely on charged-particle information.

\subsection{Track Types}
\label{sec:tracks}

After exiting the VELO a distance of $\mathcal{O}(1\,{\rm m})$ from the $pp$ collision, charged particles next traverse the first RICH detector (RICH1) before reaching a large-area silicon-strip detector located just upstream of a dipole magnet with a bending power of about $4{\rm\,Tm}$~\cite{LHCb-DP-2013-003}.
Downstream of the magnet, there are three stations of silicon-strip detectors and straw drift tubes.  
All tracking systems will be upgraded for Run~3, though only the changes to the tracking systems upstream of the magnet are relevant here.
The VELO has been redesigned to use pixels and is expected to have slightly better lifetime resolution and a lower material budget in Run~3~\cite{LHCb-TDR-013}. 
The tracking station just upstream of the magnet will also be replaced by a pixel detector and provide better coverage in $\eta$ than the current detector~\cite{LHCb-TDR-015}.  This tracking station is known as the upstream tracker (UT).  

In LHCb jargon, there are two types of tracks relevant for this search:
\begin{itemize}
\item LONG tracks that have hits in the VELO, the UT, and the stations downstream of the magnet.  These tracks have excellent momentum resolution in both magnitude and direction.
\item UP tracks that have hits in the VELO and the UT, but not in the stations downstream of the magnet.  These tracks have excellent directional resolution obtained from the VELO.   Since the curvature measurement is based only on the fringe field in which the UT operates, however, the uncertainty on the magnitude of the momentum is about 12\%~\cite{LHCb-TDR-015}.  
\end{itemize}
We also note that LHCb defines DOWN tracks which have hits in the UT and downstream of the magnet but no hits in the VELO.  While DOWN tracks are not used in this search, they could be useful for other searches involving long-lived particles.  

Charged particles may end up being reconstructed as UP tracks if they are swept out of the LHCb acceptance by the dipole magnetic field.  This may occur if a particle is produced near the edge of the detector or if it is produced with low momentum.  For simplicity, we take any charged particle with $2 < \eta < 5$ and $p > 3\gev$ to have 100\% efficiency of being reconstructed as a LONG track.  Any track that is not LONG, but satisifies  $2 < \eta < 5$ and $p > 1\gev$ is assigned as an UP track.  In reality, the reconstruction efficiency is not a step function---particles with $p < 3\gev$ may be reconstructed as LONG tracks, while particles with $p > 3\gev$ may produce UP tracks or not be reconstructed at all---but this simple choice reproduces well the overall tracking performance.  The momentum resolution for each track type is derived in \App{app:res}.  

\subsection{$D^0$ Reconstruction}
\label{subsec:dreco}

The $D^0$ meson momentum must be reconstructed for this search, since the kinematic constraints imposed by the  $D^{*0}$ mass will be used to suppress backgrounds and to improve the resolution on $m_{e^+e^-}$.   We consider two categories of $D^0$ reconstruction.  
\begin{itemize}
\item F-type: All of the $D^0$ children are charged particles so that the $D^0$ can be \emph{fully reconstructed}.  At least two of the decay products must be reconstructed as LONG tracks.  This suppresses combinatorial backgrounds and provides excellent resolution on the location of the $D^0$ decay vertex and on the $D^0$ momentum $\vec{p}_D$.   The remaining decay products are permitted to be reconstructed as either LONG or UP tracks.  
\item P-type: At least two of the $D^0$ children are reconstructed as LONG tracks so that there is excellent resolution on the location of the $D^0$ decay vertex (there may be UP tracks as well).  Requiring significant $D^0$ flight distance then permits reconstructing with good precision the direction of the $D^0$ momentum $\hat{p}_D$ using the vector from the $pp$ collision to the $D^0$ decay vertex.  For the case where the invariant mass of the missing particle(s) is known, $|\vec{p}_D|$ can be solved for as discussed below.  In this way, the $D^0$ is \emph{pseudo-fully reconstructed}.
\end{itemize}

The F-type decays considered in this search are given in \Tab{tab:dmodes}.  Each is of the form $D^0\!\to hh$ or $D^0\!\to hhhh$, where $h = K^{\pm}$ or $\pi^{\pm}$.  We do not consider doubly Cabibbo-suppressed decays (e.g.\ $D^0 \to K^+ \pi^-$) since they have small branching fractions and can be difficult experimentally to separate from the related Cabibbo-favored decays.  LHCb has already published results using most of the F-type decays listed here (see, e.g., Refs.\,\cite{Aaij:1700967,Aaij:1748270}), and each decay is expected to have minimal combinatorial background contamination even with only loose selection criteria applied.  Here we assume a baseline F-type $D^0$ selection efficiency of 90\%.  The total efficiency is then ${\rm eff}^{\rm F}_D \approx 50\%$, which is dominated by the requirement that all decay products are reconstructed by LHCb.  As shown in \App{app:res}, the resolution on $\vec{p}_D$ for F-type decays is excellent.  

\begin{table}
\begin{tabular}{cccc}
\hline \hline
Decay & $\mathcal{B}$ & $\mathcal{B}\times {\rm eff}^{\rm F}_D$ & $\mathcal{B}\times {\rm eff}^{\rm P}_D$ \\
\hline
$D^0\!\to \{K^-\pi^+,KK,\pi\pi\}$ & 4.4\% & 2.5\% & $-$ \\
$D^0\!\to \{K^-3\pi,2K2\pi,4\pi\}$ & 9.1\% & 4.5\% & 1.0\% \\
$D^0\!\to K\ell(\nu)$ & 6.8\% & $-$ &  2.0\% \\
$D^0\!\to K\pi(\pi^0)_{[0,m_{K^0}]}$ & 22.0\% & $-$ & 6.6\% \\
$D^0\!\to KK(K^0)_{[{\rm all}]}$ & 1.5\% & $-$ & 0.5\% \\
$D^0\!\to K3\pi(\pi^0)_{[0,m_{K^0}]}$ & 8.5\% & $-$ & 1.4\% \\
\hline 
Total & {} & 7.0\% & 11.5\% \\
\hline \hline
\end{tabular}
\caption{ Decays of $D^0$ mesons used in this search.  The branching fraction $\mathcal{B}$ and efficiency-corrected branching fraction are given for each decay, for both the F-type (fully reconstructed) and P-type (pseudo-fully reconstructed) selections.  The notation $(x)$ denotes that $x$ is not reconstructed.  Entries with an $[a,b]$ subscript count any decay where the invariant mass of the non-reconstructed system satisfies $a \leq m_{\rm mis} \leq b$ as signal.}
\label{tab:dmodes}
\end{table}

In P-type decays, we can use the measured flight direction to pseudo-fully reconstruct $\vec{p}_D$.  The direction $\hat{p}_D$ is a unit-normalized vector from the $pp$ collision to the $D^0$ decay vertex.  The magnitude is $|\vec{p}_D| = (\vec{p}_{\rm vis} \cdot \hat{p}_D + \vec{p}_{\rm mis} \cdot \hat{p}_D)$, where $\vec{p}_{\rm vis}$ is the reconstructed (visible) momentum and $\vec{p}_{\rm mis}$ is the non-reconstructed (missing) momentum.  Balancing the momentum transverse to the direction of flight, requires $p^{\perp}_{\rm mis} \equiv |\vec{p}_{\rm mis} \times \hat{p}_D| = |\vec{p}_{\rm vis} \times \hat{p}_D|$.
Assuming that the invariant mass of the missing decay products is known, $|\vec{p}_D|$ can be solved for in the $D^0$ rest frame using conservation of energy and the known $D^0$ meson mass.
Since $p^{\perp}_{\rm mis}$ is invariant under boosts along $\hat{p}_D$, $|\vec{p}_{\rm mis} \cdot \hat{p}_D|$ in the $D^0$ rest frame is easily obtained.  Finally, $\vec{p}_{\rm mis} \cdot \hat{p}_D$ can be determined in the lab frame up to a two-fold ambiguity that arises because the sign of  $\vec{p}_{\rm mis} \cdot \hat{p}_D$ in the $D^0$ rest frame is not known.  However, once the $D^0$ is combined with an \Adecay candidate to form a $D^{*0}$ candidate, the vast majority of the time only the correct solution produces an invariant mass consistent with that of the $D^{*0}$ meson.  
As described in \App{app:res}, we take the baseline selection efficiency for P-type decays to be 50\%, since the $D^0$ flight distance must be large relative to the vertex resolution to obtain good resolution on $\vec{p}_D$.

The P-type decays considered in this search are given in \Tab{tab:dmodes}.  We again do not consider doubly Cabibbo-suppressed decays.  Other decays that are ignored include those where the missing mass cannot be reliably predicted, such as $D^0\!\to \pi^-\ell^+(X)$, which dominantly has $X=K^0\nu$ as the missing system.
Note that solving for $|\vec{p}_D|$ in P-type decays requires using the known missing mass as a constraint.  That said, the resolution is only degraded slightly if the true missing mass differs from that used in the reconstruction by up to about $0.2m_{D^0}$.  For example, when the visible part of the decay is $K^-\pi^+$, the most likely missing system is a single $\pi^0$; if the missing mass is taken to be $m_{\pi^0}$, but the actual decay is $D^0\!\to K^-\pi^+ K^0$, the resolution obtained on $m_{e^+e^-}$ by applying the ``wrong'' kinematical constraints to the $D^{*0}$ candidate is only worse by about 10\%.  In \Tab{tab:dmodes}, we list the missing mass ranges considered as signal for each P-type decay.  Candidates where the missing mass falls outside of these windows are ignored in this analysis, since they have worse resolution and anyways make up a small fraction of the P-type decays.
A derivation of the P-type $D^0$ resolution is given in \App{app:res}.  The resolution on $\vec{p}_D$ is about an order of magnitude worse in P-type than F-type decays; however, the $m_{\A}$ resolution after performing a mass-constrained fit is similar (as shown in \Fig{fig:mee} below).

\subsection{$D^{*0}$ Reconstruction}
\label{subsec:dstarreco}

To reduce the background from unassociated $D^0 e^+ e^-$ combinations, we require that the reconstructed mass difference
\be
\Delta m^{\rm reco}_D = m_{\rm reco}(D^0 e^+ e^-) - m_{\rm reco}(D^0)
\ee
satisfies
\be
\label{eq:deltamrecorequirement}
-50 \mev < \Delta m^{\rm reco}_D - \Delta m_D < 20 \mev.
\ee
The looser requirement is placed on the lower edge due to bremsstrahlung by the electrons.    This mass requirement highly suppresses the decay $D^{*0}\!\to D^0 \pi^0(\gamma e^+ e^-)$ and its \A counterpart, except when \mee is large (see \Sec{subsec:piondiscussion} below).  The efficiency of this requirement is about ${\rm eff}_{\Delta m_D} \approx 85\%$.  Note that this cut can be tightened at the expense of signal efficiency if combinatorial backgrounds turn out to be problematic (see \Sec{subsec:additionalbackgrounds} below).

\subsection{\A Reconstruction}
\label{subsec:aprimereco}

The reconstructed electrons produced in $\A\!\to e^+e^-$ decays are a mixture of UP and LONG tracks. Only a few percent of the electrons have momenta large enough that equivalent-momenta non-electrons would be able to emit Cherenkov light in RICH1.  Therefore, identification of the $e^+$ and $e^-$ should be highly efficient with a low hadron-misidentification rate.  Furthermore, the signature of a maximum-Cherenkov-angle ring in coincidence with a track should suppress the fake-track background which can be sizable at low momenta.

Bremsstrahlung radiation and multiple scattering of the electrons significantly affect the \mee resolution.  We implement this numerically in our simulation following Refs.~\cite{PDG2014,Koch:1959zz} and using the Run~3 LHCb VELO~\cite{LHCb-TDR-013}, RICH1~\cite{LHCb-TDR-014}, and UT~\cite{LHCb-TDR-015} material budgets.  Bremsstrahlung downstream of the magnet does not affect the momentum measurement and is ignored.

\begin{figure}[t]
\includegraphics[width=0.96\columnwidth]{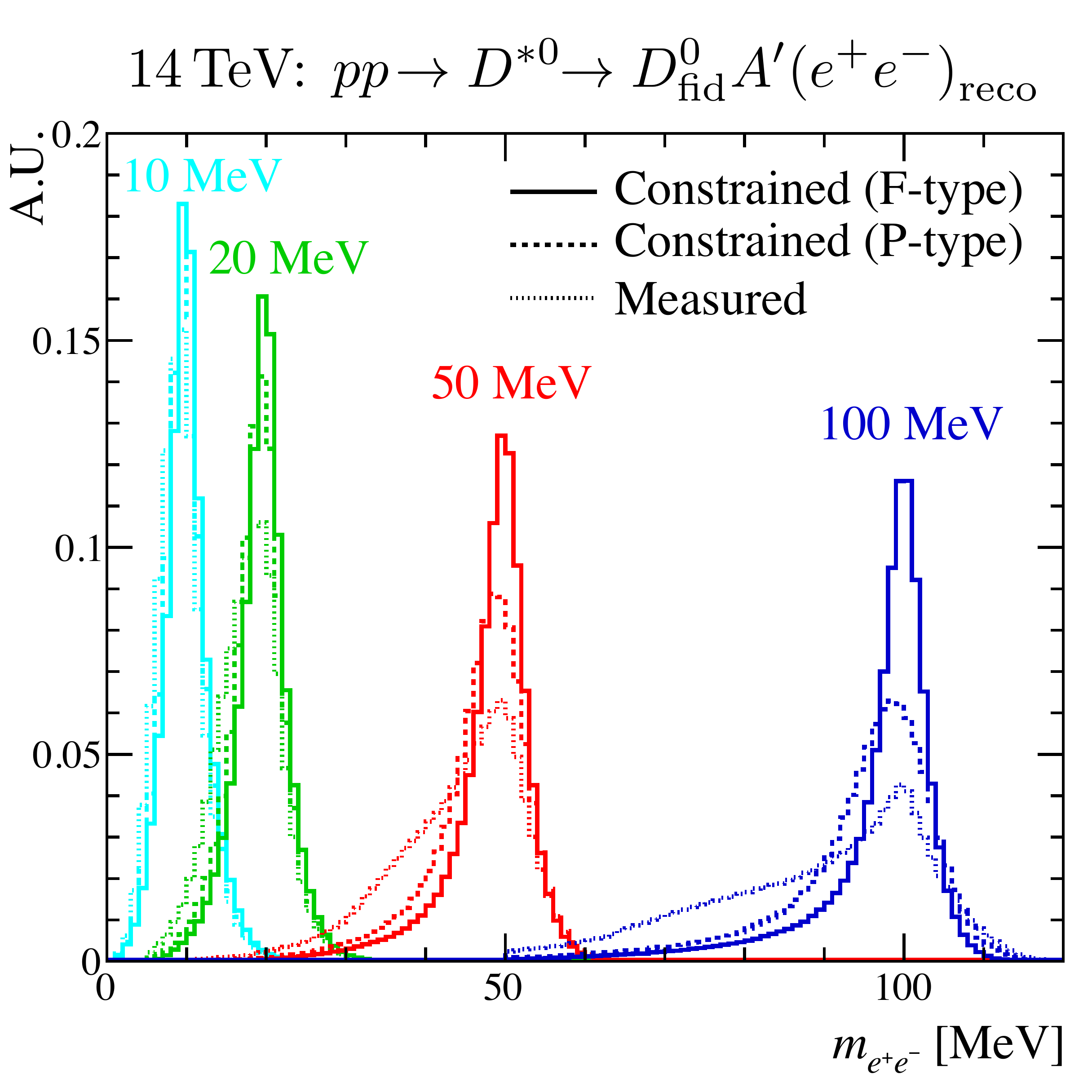}
\caption{Distribution of \mee with (solid, dashed) and without (dotted) incorporating the $D^{*0}$ mass constraint for $m_{\A} = \{10,20,50,100\}\mev$.  The solid curve shows better performance than the dashed one because F-type $D^0$ candidates have better momentum resolution than P-type ones.}
\label{fig:mee}
\end{figure}

In \Fig{fig:mee}, we show the resolution on \mee for several values of $m_{\A}$, where the \A candidates are constrained to originate from the $pp$ collision.  Bremsstrahlung creates large low-mass tails resulting in poor resolution on \mee.  Since the $D^{*0}$ mass is known and its width is less than the detector resolution, though, we can correct the \mee distribution once we identify the $D^{*0}$ candidate and apply the $\Delta m^{\rm reco}_D$ cut.  As a heuristic, one can rescale the \mee value by a simple correction factor
\be
\label{eq:meecor}
m_{\epem}^{\rm corr} =  m_{\epem}^{\rm reco}\left(2 - \frac{\Delta m_D^{\rm reco}}{\Delta m_D}\right).
\ee
A more sophisticated approach involves performing a mass-constrained fit to enforce energy-momentum conservation and the known $D^{*0}$ mass using the covariance matrices of all reconstructed particles.  Using this fit, we find 10--20\% improvement in $\sigma(\mee)$ relative to the simple correction given in \Eq{eq:meecor}.   
As shown in \Fig{fig:mee}, the resolution on \mee after the applying the kinematic fit is 2--3\mev using F-type $D^0$ candidates, and 2--5\mev using P-type $D^0$ candidates.

The key difference between the pre-module displaced, post-module displaced, and resonant searches are the requirements placed on the $\A$ flight distance.  These are described in more detail in the subsequent sections.

\section{Displaced \A Search (Pre-Module)}
\label{sec:reach_dis}

The \A typically has a large Lorentz boost factor, resulting in the \A decay vertex being significantly displaced from the $pp$ collision for $\epsilon^2 \lesssim 10^{-7}$.  The combined signature of a displaced $D^0$ decay vertex, a displaced \Adecay vertex, $m(D^0\A)$ consistent with $m(D^{*0})$, and a consistent decay topology will result in a nearly background-free search.  This pre-module displaced search is aimed at $\A$ decay vertices that occur within the beam vacuum upstream of the first VELO module intersected by the \A trajectory.

\subsection{Conversion and Misreconstruction Backgrounds}
\label{subsec:natalia}

At LHCb, the first layer of material is the foil that separates the beam vacuum from the VELO vacuum.  This foil is corrugated to accommodate the VELO modules, such that if the $\A$ decays prior to the foil, it still effectively decays within the VELO tracking volume.  The average transverse distance that the $\A$ will travel before hitting a VELO module is $6~\mm$ \cite{LHCb-TDR-013}, which, because of the corrugated foil geometry, is roughly the average transverse flight distance to the foil as well.

To effectively eliminate backgrounds from $\gamma \to e^+ e^-$ conversions in the foil, we require the $\A$ decay vertex to be reconstructed upstream of the foil.  Furthermore, each reconstructed electron must have an associated hit in the first relevant VELO module given the location of the reconstructed \A decay vertex.  These hits are required to have at least one vacant VELO pixel between them to avoid any charge-sharing issues, imposing an effective buffer distance between the \A decay vertex and the foil:
\be
\label{eq:Dbuffer}
D \approx \frac{0.123\,\mm}{\alpha_{e^+e^-}},
\ee
where $\alpha_{e^+e^-}$ is the electron-positron opening angle.  In reality, the VELO pixels in Run~3 will be $55\times55\,\mu{\rm m}^2$ squares; the definition of $D$ is based on treating the pixels as circles with 0.123\,mm being twice the effective diameter (the precise value used here has no impact on our search).  
The pre-module $\A$ requirement can then be approximated by requiring the $\A$ transverse flight distance to satisfy
\be
\label{eq:prefoilrequirement}
\ell_{\rm T} < 6~\mm - D_{\rm T}, \quad D_{\rm T} = D \sin \theta,
\ee
where $\theta$ gives the $\A$ flight direction.  To remove \A trajectories that first intersect the foil far from a module, we require $\eta_\A > 2.6$.  We also impose $\eta_\A < 5$ to avoid possible contamination due to $pp$ collisions that are not properly reconstructed.\footnote{An \A candidate may be accidentally formed from a prompt $e^+e^-$ pair produced in a $pp$ collision if the event is not properly reconstructed.  In particular, if a $D^0$ meson is produced in another $pp$ collision upstream of that interaction point, the ``displaced'' $\A$ would produce a consistent decay topology, albeit with $\eta_\A \to \infty$.}

Having suppressed conversion backgrounds, the dominant background comes from prompt $D^{*0} \to D^0 e^+ e^-$  events where the $e^+e^-$ vertex is misreconstructed as being displaced because of multiple scattering of the electrons in the detector material.  We estimate this background in a toy simulation of the Run~3 VELO, taking scattering angle distributions from a \geant simulation  which includes non-Gaussian Moli\`{e}re scattering tails.\footnote{It is likely that \geant overestimates the probability for large-angle scatterings (see Ref.\,\cite{HPS2014}).  If so, our results are conservative, since these scattering tails effectively define the reach for the pre-module $\A$ search.}  Many of these fake $\A$ vertices can be eliminated by requiring a consistent decay topology, in particular that the angle between $\vec{p}_{\A}$ and the vector formed from the $pp$ collision to the \A decay vertex is consistent with zero, and the electrons travel within a consistent decay plane.

The remaining misreconstructed background events have a consistent topology, so a cut on transverse flight distance $\ell_{\rm T}$ is required to ensure a significant displaced $\A$ vertex.  To avoid fake displaced vertices from one electron experiencing a large-angle scattering, we also require both the electron and positron to have a non-trivial impact parameter (IP) with respect to the $pp$ collision.  These requirements are summarized by
\be
\label{eq:prefoildisplaced}
\ell_{\rm T} > n \, \sigma_{\rm \ell_{\rm T}}, \quad {\rm IP}_{e^\pm} > \frac{n}{2} \sigma_{\rm IP},
\ee
where the value of $n$ is adjusted to yield $\approx 1$ background event in each $\A$ mass window, with $n$ ranging from 3 to 5 as a function of $m_{A'}$.  The selection in \Eq{eq:prefoildisplaced} is meant to be simple and robust, and could certainly be optimized in a full analysis.  See \App{app:vertex} for details on the $\ell_{\rm T}$ and IP resolution.

\subsection{Event Selection}
\label{subsec:eventselection}

\begin{figure}[t]
\includegraphics[width=0.96\columnwidth]{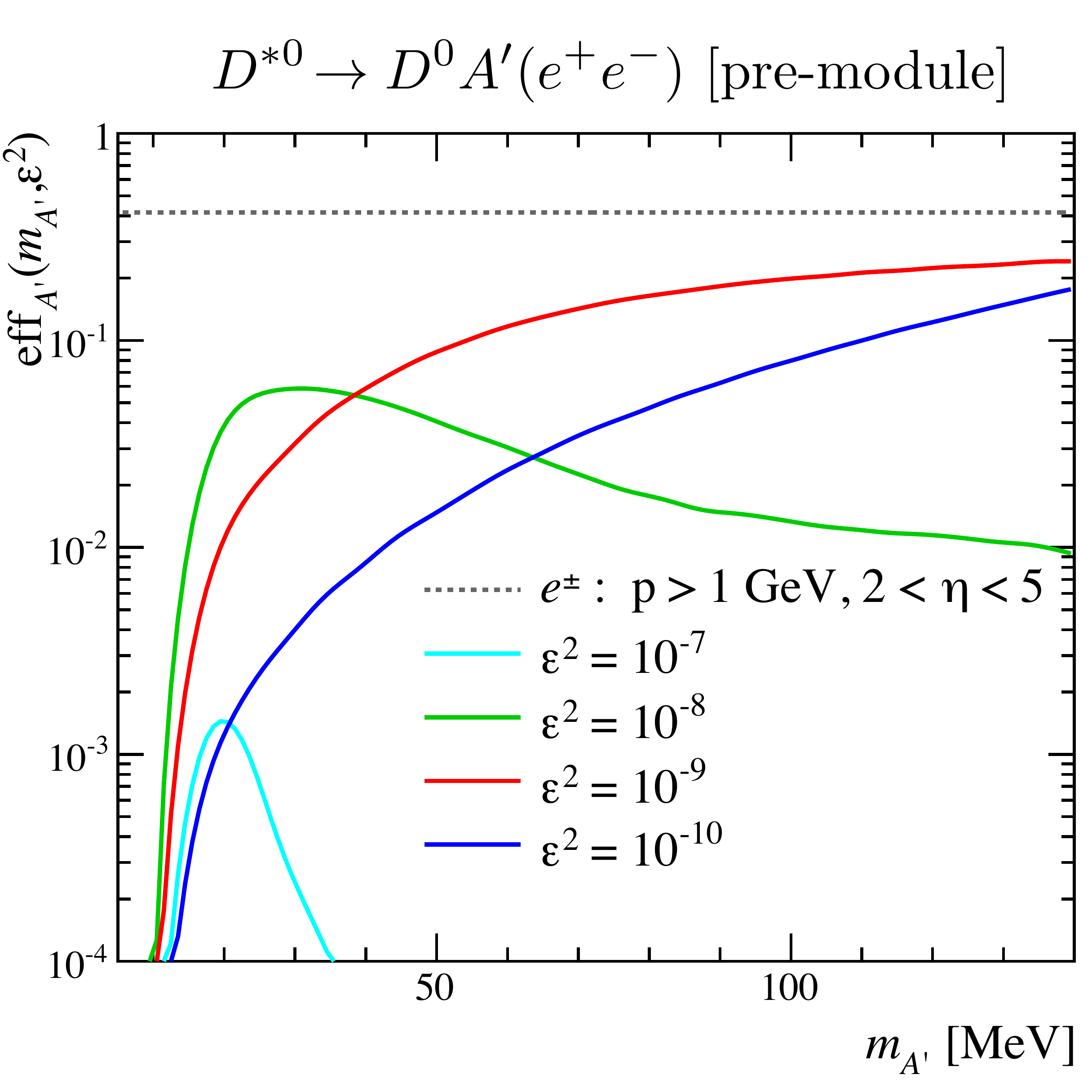}
\caption{Pre-module displaced search \A signal efficiency ${\rm eff}_{\A}(m_{\A},\epsilon^2)$ for several $\epsilon^2$ values.  The dashed line shows the efficiency of the track-reconstruction requirements placed on the electrons alone, while the solid lines show the total \A efficiency.  
This efficiency does not include the contribution from the $D^0$ or $D^{*0}$ selection (i.e.\ ${\rm eff}^{\rm F,P}_D$ or  ${\rm eff}_{\Delta m_D}$).  
}  
\label{fig:dispsigacc}
\end{figure}

Summarizing, the event selection for the pre-module displaced $A'$ search is:
\begin{itemize}
  \item F-type or P-type $D^0$ candidate;
  \item $e^+$ and $e^-$ from LONG or UP tracks with hits in the first VELO module they intersect;
  \item reconstructed $D^{*0}$ candidate from the $D^0$, $e^+$, and $e^-$;
  \item reconstructed $A' \to e^+ e^-$ satisfying the conversion veto ($\ell_{\rm T} < 6~\mm - D_{\rm T}$, $\eta_{A'} \in [2.6,5]$);
  \item reconstructed $A' \to e^+ e^-$ with significant displacement ($\ell_{\rm T} > n \, \sigma_{\rm \ell_{\rm T}}, {\rm IP}_{e^\pm} > \frac{n}{2} \sigma_{\rm IP}$).
\end{itemize}
In \Fig{fig:dispsigacc}, we show the resulting \A signal efficiency ${\rm eff}_{\A}(m_{\A},\epsilon^2)$. For $\epsilon^2 = 10^{-10}$, which is near the low end of the reach, the efficiency is limited by the efficiency of the conversion veto.  As $\epsilon$ increases, the requirement of a significant displacement ultimately limits the reach of the displaced search.

\subsection{Additional Backgrounds}
\label{subsec:additionalbackgrounds}

Beyond the misreconstructed $D^{*0} \to D^0 e^+ e^-$ background, a full accounting of the potential backgrounds for the displaced search is difficult since all SM processes with large rates are highly suppressed.  Therefore, any additional backgrounds will be dominated by extremely rare processes or highly unlikely coincidences.

One possible source of backgrounds would be $B\!\to D^{*0}(D^0\epem) X$ decays, since the resulting \epem vertex is truly displaced.  Such decays can be suppressed by making the following requirements: the $D^0$ and \A momenta must intersect the $pp$-collision point when traced upstream from their respective decay vertices; the $D^{*0}$ decay vertex is consistent with the $pp$-collision vertex; and there are no additional tracks consistent with originating from the \A decay vertex.  Furthermore, one could require that the \A decay vertex is downstream of the $D^0$ decay vertex, which would be efficient for the smaller $\epsilon$ values probed in this search.  Therefore, we do not expect a significant amount of background coming from $B\!\to D^{*0}(D^0\epem) X$ decays.  

The decays of other long-lived mesons could also be sources of displaced \epem vertices.  Decays of charged pions and kaons that produce an \epem pair are rare, though, and the probability for these particles to decay in the VELO is small.  A more likely source is the decay $\pi^0\!\to\epem\gamma$, where the $\pi^0$ is produced in the decay of a long-lived meson.  All of the other meson-decay products must be neutral, of course, otherwise the presence of additional charged particles consistent with originating from the \epem vertex could be used as a veto.  For example, the decays $K_S\!\to\pi^0\pi^0$ and $D^0\!\to K_S\pi^0$ occur with huge rates within the LHCb VELO.  Such decays, however, are unlikely to result in the \A candidate momentum intersecting the $pp$-collision point or to occur in coincidence with a $D^0$ meson such that $m(D^0\A)$ is consistent with $m(D^{*0})$.  

To see whether we could estimate displaced combinatorial backgrounds in Monte Carlo, we generated a sample of 30 million \pythia $pp$ collisions at 14\tev.  We found that no combination of a true $D^0$ with two displaced tracks (not necessarily electrons, but assigned the electron mass) had an invariant mass within the $D^{*0}$ mass window.  In this simulated sample, there are only three candidates with $m(D^0e^+e^-)-m(D^0) < 500\mev$ using true electrons and none within 150\mev of our $D^{*0}$ mass window.  However, we are anticipating $10^8$ times more $D^{*0}$ meson decays in the full LHCb data sample, so it appears that it is not feasible to use Monte Carlo to precisely estimate the displaced combinatorial background.  This type of background will therefore need to be examined in data using the $\Delta m_D^{\rm reco}$ sidebands.  If specific sources of combinatorial background are identified as problematic, then the selection will need to be adjusted to remove them.

\subsection{Contribution from Pion Decays}
\label{subsec:piondiscussion}

Thus far, we have ignored the channel $D^{*0}\!\to D^0\pi^0(\gamma \A)$, which is another potential source of signal events.  Decays of this type are highly suppressed by the $\Delta m_D^{\rm reco}$ requirement in \Eq{eq:deltamrecorequirement}, though, unless $m_{\A}$ is large.  For most allowed $m_{\A}$ values, one can choose whether or not to include such decays in the analysis by adjusting the $\Delta m_D^{\rm reco}$ requirement.  After removing the $\Delta m_D^{\rm reco}$ requirement, the expected yields of $D^{*0}\!\to D^0\pi^0(\gamma \A)$ and \sigdecay decays are comparable, but so is the expected background contamination from misreconstructed $D^{*0}\!\to D^0\pi^0(\gamma e^+ e^-)$ and $D^{*0}\!\to D^0 e^+ e^-$.  We choose not to include this channel when estimating the reach below, but note that such decays may prove useful in a complete analysis.
 
If one does try to use the $D^{*0}\!\to D^0\pi^0(\gamma \A)$ channel, then one should be aware of an important subtlety when incorporating \mee information.  As described in \Sec{subsec:aprimereco}, a kinematic fit can be used to improve the \mee resolution.   For $D^{*0}\!\to D^0\pi^0(\A\gamma)$ decays, however, the missing $\gamma$ is not accounted for when enforcing energy-momentum conservation.   We find that this results in the $m_{\A}$ peak being shifted up in mass by about 20\mev, with the resolution on \mee degraded by about a factor of two.  This results in two peaks in the reconstructed $m_{\A}$ spectrum, coming from the $D^{*0}\!\to D^0 \A$ and $D^{*0}\!\to D^0\pi^0(\gamma \A)$ channels.  If the background level is low, then this second peak could be used to boost the significance of an \A signal.  Indeed, one narrow peak with a second, wider peak shifted in mass by a fixed amount would be a striking signature.  If the background level is high, then this wider peak would largely be absorbed into the background and have no impact on the signal significance.  

Finally, photon conversions arising from $D^{*0}\!\to D^0\pi^0(\gamma\gamma)$ decays are also highly suppressed by the $\Delta m_D^{\rm reco}$ requirement, and in the absence of misreconstruction, can be eliminated by the pre-module requirement in \Eq{eq:prefoilrequirement}.  We expect such conversions to contribute less than those from \smdecay decays.   

\subsection{Reach}
\label{subsec:reach_dis}

The expected signal yield for the displaced search as a function of $m_{\A}$ and $\epsilon^2$ is given by
\begin{eqnarray}
S(m_{\A},\epsilon^2) &=& N(\smdecay)\, \frac{\Gamma(\sigdecay)}{\Gamma(\smdecay)} \, {\rm eff}_{\Delta m_D}   \nonumber \\
&& \qquad \times \left({\rm eff}^{\rm F}_D + {\rm eff}^{\rm P}_D \right) {\rm eff}_{\A}(m_{\A},\epsilon^2) \nonumber \\
&& \hspace{-0.7in} \simeq 85  \left( \frac{\epsilon^2}{10^{-10}} \right) \left(1 - \frac{m_{\A}^2}{\Delta m_D^2} \right)^{3/2} {\rm eff}_{\A}(m_{\A},\epsilon^2).  \label{eq:displaced_reach}
\end{eqnarray}
As discussed above, we adjusted the requirement in \Eq{eq:prefoildisplaced} to ensure $\approx 1$ background event in any given $m_{A'}$ window.   
Assuming that all relevant backgrounds have been accounted for, the reach would be set at 95\% confidence level by requiring $S \gtrsim 3$.  However, to allow for the possibility of extremely rare background sources,
the reach is set by requiring $S \ge 5$.  In this way, we account for either additional background candidates in the final data sample or for a lower \A selection efficiency due to the criteria required to suppress these additional backgrounds.
The reach is shown in \Fig{fig:lhcbbounds} assuming 15\,fb$^{-1}$ of data collected by LHCb in Run~3, which covers a significant part of the allowed parameter space for $m_{\A} \lesssim 100$\:MeV.  

\section{Displaced \A Search (Post-Module)}
\label{sec:reach_dis_post}

In order to capture more $A'$ signal events, one can effectively reverse the pre-module requirements in \Eq{eq:prefoilrequirement} and search for post-module $\A$ decays.  Here, the dominant background is $D^{*0}\!\to D^0\gamma$, where the on-shell $\gamma$ converts into $e^+ e^-$ via interactions with the detector material.  As we will see, this post-module search does not cover much additional $\A$ parameter space compared to the pre-module search, but is important as a cross check of a possible \A discovery.

\subsection{Misreconstruction and Conversion Backgrounds}
\label{subsec:convveto}

The background considerations in the post-module case are reversed compared to the pre-module case in \Sec{subsec:natalia}.  Here, the background from misreconstructed $D^{*0} \to D^0 e^+ e^-$ events can be effectively eliminated by requiring \emph{no} hits in the first VELO module intersected by the reconstructed electron trajectories.  

The dominant background in the post-module search comes from \smdecay with photon conversions.  We simulate this background using the Run~3 LHCb VELO material as described in \Ref{LHCb-TDR-013} with the Bethe-Heitler \mee spectrum as given in \Ref{PhysRev.89.1023}.\footnote{It is vital that all \A searches use the Bethe-Heitler spectrum, rather than the one produced by \geant.  \geant vastly underestimates the fraction of conversions that produce large \mee due to the usage of a less-CPU-intensive approximation of the Bethe-Heitler equation.}  We start with electron tracks that each have at least three hits in the VELO.  This imposes an effective \A transverse flight distance requirement of
\be
\ell_{\rm T} \in [6\, \mm, 22 \, \mm].
\ee
In reality, the electron hit requirement does not result in a step function for the \A efficiency. However, in the long-lifetime limit, this simple approximation produces the same integrated efficiency.
We then require the reconstructed $\A$ vertex to be significantly displaced from the VELO material.  This can be well-approximated by treating the VELO as a stack of modules located at longitudinal distances $z_i = i \cdot 25~\mm$, where $z$ is measured from the point where the $\A$ has $\ell_{\rm T} = 6~\mm$ (i.e.~the average position where the $\A$ trajectory crosses a VELO module).  From a given $\A$ decay vertex at a location $z$ between modules $i$ and $i+1$, one requires
\be
z- z_i > n \, \sigma_z, \quad z_{i+1} - z > D, \quad {\rm IP}_{i,e^{\pm}} > \frac{n}{2}\sigma_{{\rm IP}_i},
\ee
where $D$ is the same buffer distance in \Eq{eq:Dbuffer} and IP$_i$ is defined with respect to the location where the $A'$ trajectory intersects the $i$-th module (see \App{app:vertex} for the corresponding resolutions).  We also impose the same $\eta_{A'} \in [2.6,5]$ requirement as in \Sec{subsec:natalia}.  Using our simulation, we adjust $n$ such that $\approx 1$ $D^{*0}\!\to D^0\gamma,\gamma\!\to\epem$ event will survive these criteria in each $\A$ mass window.

\subsection{Event Selection and Reach}

\begin{figure}[t]
\includegraphics[width=0.96\columnwidth]{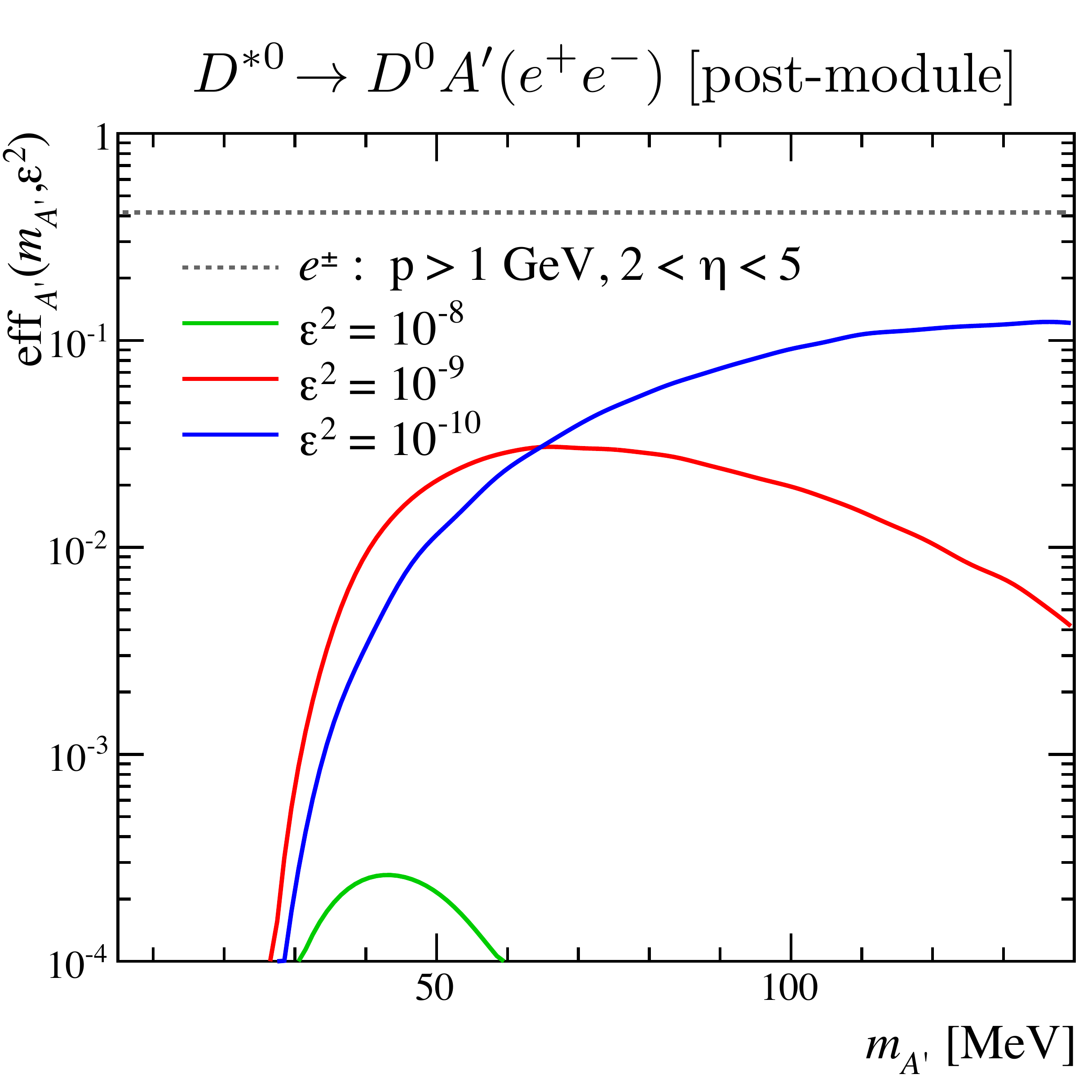}
\caption{Post-module displaced search \A signal efficiency ${\rm eff}_{\A}(m_{\A},\epsilon^2)$ for several $\epsilon^2$ values.  The dashed line shows the efficiency of the track-reconstruction requirements placed on the electrons alone, while the solid lines show the total \A efficiency.  
This efficiency does not include the contribution from the $D^0$ or $D^{*0}$ selection (i.e.\ ${\rm eff}^{\rm F,P}_D$ or  ${\rm eff}_{\Delta m_D}$).  
}  
\label{fig:dispsigacc_post}
\end{figure}

Summarizing, the event selection for the post-module displaced $A'$ search is:
\begin{itemize}
  \item F-type or P-type $D^0$ candidate;
  \item $e^+$ and $e^-$ from LONG or UP tracks with no hits in the first VELO module they intersect;
  \item reconstructed $D^{*0}$ candidate from the $D^0$, $e^+$, and $e^-$;
  \item reconstructed $A' \to e^+ e^-$ satisfying the prompt veto ($\ell_{\rm T} \in [6\, \mm, 22 \, \mm]$, $\eta_{A'} \in [2.6,5]$);
  \item reconstructed $A' \to e^+ e^-$ with significant displacement from VELO modules ($\Delta z_i > n \, \sigma_z$, ${\rm IP}_{i,e^\pm} > \frac{n}{2} \sigma_{{\rm IP}_i}$, $\Delta z_{i+1} > D$).
\end{itemize}
In \Fig{fig:dispsigacc_post}, we show the resulting \A signal efficiency ${\rm eff}_{\A}(m_{\A},\epsilon^2)$. 

The formula for the reach for the post-module displaced search is identical to \Eq{eq:displaced_reach}, albeit with a modified signal efficiency.  
We again set the reach using $S \ge 5$ to allow for rare, unaccounted for background sources.
As shown in \Fig{fig:lhcbbounds}, the reach in the post-module search is not any better than the pre-module search.  The reason is that for small enough $\epsilon^2$, the decay probability as a function of $\ell_{\rm T}$ is constant, and there is comparable efficiency for the $\A$ decay vertex to be upstream or downstream of the first VELO module.  As $\epsilon^2$ increases, the post-module $\ell_{\rm T} > 6\, \mm$ requirement becomes inefficient before the pre-module vertex requirements do, giving the pre-module search a better reach.  Of course, a slightly better limit could be set by combining the pre-module and post-module requirements, but we do not do this here since the dominant background sources are different.  We also note that if a discovery is made, the ability to confirm the presence of a signal in both displaced regions will provide a powerful systematic check. 

\subsection{Additional Backgrounds}
\label{subsec:additionalbackgrounds_post}

The same additional backgrounds from \Sec{subsec:additionalbackgrounds} might affect the post-module search, with the exception of $\pi^0$ mesons produced in charm-meson decays.  In addition, there is a potential background from improper reconstruction of photon conversion events.

One way to misreconstruct a $\gamma$ conversion as an \A decay is if the $\gamma$ converts in or just in front of a VELO module.  
Since the separation between the $e^+$ and $e^-$ when they traverse the VELO module would be less than the hit resolution, only a single hit would be recorded.
The positron track may be formed using this hit.  The electron track would be missing a hit in this module and fail the selection requirements, unless an unassociated hit happens to occur close by.  
The hit occupancy expected in the VELO during Run~3 is about 0.08\% in the inner-most pixels, and less than 0.01\% for pixels more than 10\,mm from the beam line~\cite{LHCb-TDR-013}.  
Since we require a full vacant pixel between the first $e^+$ and $e^-$ hits, there must be a large angle between the electron momentum and the vector $\vec{h}$ formed from the unassociated hit to the $e^-$ hit in the second module.
Therefore, a consistent track will only be formed if the electron undergoes an unlikely scatter in the second module so that the hit in the third module is consistent with $\vec{h}$.  
We cannot reliably estimate the probability of this coincidence in our toy simulation, but expect it to be at worst comparable to the remaining conversion backgrounds left after the requirements in \Sec{subsec:convveto}, so it should not affect the predicted reach.

\section{Resonant Search} 
\label{sec:reach_res}

When $\epsilon^2$ is large enough, the $\A$ decay vertex is no longer significantly displaced from the $pp$ collision.  In this case, we have to rely on reconstructing an \A mass peak.  This search is also relevant for non-minimal scenarios with a larger dark photon width, such as when the dark photon has an invisible decay to dark matter particles (see, e.g., \Refs{Aubert:2008as,Batell:2009di,deNiverville:2011it,Wojtsekhowski:2012zq,Kahn:2012br,Izaguirre:2013uxa,Batell:2014mga,Kahn:2014sra,Izaguirre:2014bca,Izaguirre:2015yja}).  

For simplicity, we perform our analysis still assuming the minimal $\Gamma_{\A}$ value in \Eq{eq:gamma_a}.  This means that conversion backgrounds are important, so we apply the conversion veto from \Sec{subsec:natalia}.  The dominant background becomes prompt $D^{*0}\!\to D^0\epem$, which is irreducible, and the resonant reach is limited by the \mee resolution shown in \Fig{fig:mee}.  To ensure good $m_{\epem}$ resolution, we further impose an \A opening angle cut $\alpha_{\epem} > 3$\,mrad.

\begin{figure}[t]
\includegraphics[width=1.0\columnwidth]{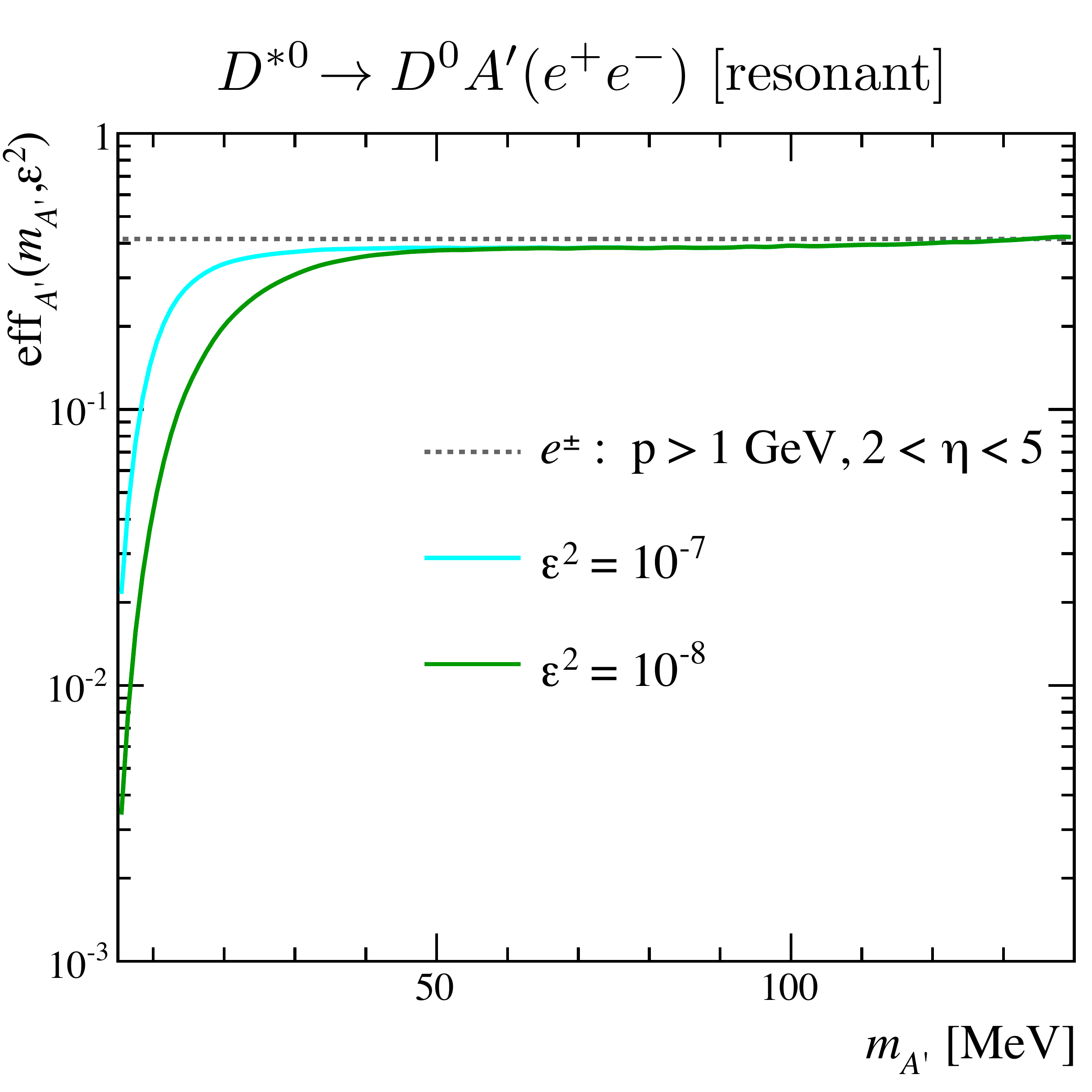}
\caption{
 Resonant search \A signal efficiency ${\rm eff}_{\A}(m_{\A},\epsilon^2)$.  The dashed line shows the efficiency to reconstruct both electrons, while the solid line shows the total \A efficiency.  The efficiency is nearly independent of $\epsilon$ for $\epsilon^2 > 10^{-7}$.
This efficiency does not include the contribution from the $D^0$ or $D^{*0}$ selection (i.e.\ ${\rm eff}^{\rm F,P}_D$ or  ${\rm eff}_{\Delta m_D}$).
}
\label{fig:dispresacc}
\end{figure}

In summary, the event selection criteria for the resonant search are
\begin{itemize}
  \item F-type or P-type $D^0$ candidate;
  \item $e^+$ and $e^-$ from LONG or UP tracks with hits in the first VELO module they intersect;
    \item reconstructed $D^{*0}$ candidate from the $D^0$, $e^+$, and $e^-$;
    \item reconstructed $A' \to e^+ e^-$ satisfying conversion veto ($\ell_{\rm T} < 6~\mm - D_{\rm T}$, $\eta_{A'} \in [2.6,5]$);
  \item  \A decay opening angle $> 3$\,mrad to ensure good $m_{\epem}$ resolution.
\end{itemize}
In \Fig{fig:dispresacc}, we show the \A signal efficiency ${\rm eff}_{\A}(m_{\A},\epsilon^2)$ for the resonant search.  Apart from the electron track-reconstruction, the dominant source of inefficiency is the opening angle requirement.

Due to the large irreducible background level in this search, we assume that $D^{*0}\!\to D^0\pi^0(\gamma\A)$ decays provide negligible additional sensitivity and therefore ignore them.  Background contamination from $D^{*0}\!\to D^0\pi^0(\gamma\epem)$ decays is still included, but it is highly suppressed by the $\Delta m_D^{\rm reco}$ requirement except when \mee is large.  

To determine the reach in the resonant search, we require $S/\sqrt{B} > 2$ to obtain a 95\% confidence limit.  The signal significance is given by
\begin{eqnarray}
\label{eq:soversqrtb}
\frac{S}{\sqrt{B}} = \Gamma(\sigdecay)\sqrt{\frac{N\left(D^{*0}\!\to D^0\gamma \right)}{\Gamma(D^{*0}\!\to D^0\gamma)\Delta\Gamma}} \hspace{0.2in} \nonumber \\ 
\times \sqrt{ {\rm eff}_{\Delta m_D} \left({\rm eff}^{\rm F}_D+{\rm eff}^{\rm P}_D\right) {\rm eff}_{\A}(m_{\A},\epsilon^2)},
\end{eqnarray}
where
\be
\label{eq:resonantmasswindow}
\Delta\Gamma \equiv \int_{m_{\A}-\Delta m_{\A}}^{m_{\A}+\Delta m_{\A}} {\rm d}\mee \frac{{\rm d}\Gamma(D^{*0}\!\to D^0\epem)}{{\rm d}\mee}
\ee
and $\Delta m_{\A} = 2\,\sigma(\mee)$ is evaluated at $m_{\A}$.\footnote{Because F-type and P-type $D^0$ mesons yield different \A mass resolutions, \Eq{eq:resonantmasswindow} is evaluated using the appropriately weighted average of the two event categories.}  Here, we are assuming that ${\rm eff}_{\A}(m_{\A},\epsilon^2)$ is constant over $[m_{\A}-\Delta m_{\A},m_{\A}+\Delta m_{\A}]$.   The reach for the resonant search is shown in \Fig{fig:lhcbbounds}, and combined with the pre-module displaced search, it closes most of the available \A parameter space for $m_\A \lesssim 100$\:MeV.

The triggerless-readout system, along with real-time data calibration, will make it possible to identify \sigdecay candidates online during data taking.  For the displaced searches, the number of candidates recorded for further analysis can be made as small as required since the dominant backgrounds are reducible.  The resonant search, on the other hand, relies on looking for an \mee peak in $\mathcal{O}(10^9)$ $D^{*0}\!\to D^0\epem$ decays.  Recording billions of full events for such a search is simply not feasible.   That said, LHCb is already commissioning in Run~2 partial event storage for use in studying high-rate processes~\cite{LHCb-PAPER-2015-041}.  This involves storing only the information relevant to the signal candidate and discarding the rest of the event information. 

In terms of required bandwidth, in principle the resonant search can be carried out by storing just \mee for each signal candidate, though in practice one wants to store additional information to perform cross checks.  At minimum, we expect that the four-momenta and various detector-response information can be kept for all particles in each signal candidate.  More information can be kept for displaced candidates that are inconsistent with arising due to photon conversion, where the criteria used to select such candidates will be much looser than those applied to define our reach above.  Finally, full events can be kept for a small fraction of signal candidates to permit detailed offline studies of the detector performance and background contributions.   The experience gained by LHCb during Run~2 using reduced event storage will aid in determining how to optimize the data-storage strategy employed for the \sigdecay search.  

\section{Possible Improvements}
\label{sec:improve}

The predicted reach in \Fig{fig:lhcbbounds} covers most of the available \A parameter space for $m_\A \lesssim 100$\;MeV, and one expects that LHCb could fully cover this region with modest improvements.   That said, this reach is based on simulation, so despite our conservative approach, the actual reach may be less than predicted.  Furthermore, the ultimate goal is to discover an \A boson and measure its properties.  Discovery requires $5\sigma$ sensitivity, which our search provides over most of the relevant \A parameter space but not all of it.  Precision measurement of the \A mass and lifetime may be vital to determining whether the \A decays invisibly into dark matter.   For these reasons, it is worth investigating possible improvements.  

There are various ways to improve the dark photon search at LHCb.  At minimum, one could assume that the integrated luminosity accumulated by LHCb is larger than the 15\:fb$^{-1}$ baseline used in this study.\footnote{As stated in footnote \ref{footnote:charmxsec}, the first measurement of the $c\bar{c}$ cross section at 13\tev suggests that the cross section in \Eq{eq:baselinexsec} is likely about 20\% too small.  If so, one can view our reach results as for 12\,fb$^{-1}$.}  
One year of running at the nominal instantaneous luminosity corresponds to about 10\,fb$^{-1}$; therefore, our Run~3 estimate conservatively provides substantial LHC ramp up and LHCb commissioning time.  LHCb also plans to take data in Run~4, collecting a minimum of 50\,fb$^{-1}$ in Runs 3 and 4 combined~\cite{LHCb-TDR-016}.  Furthermore, LHCb is currently investigating ways to increase the integrated luminosity in Run~4.  By the end of Run~4, LHCb may collect 10--20 times more data than we used to estimate the reach in this study.

There are several ways to improve the \mee resolution.  The mass-constrained fit we employed in \Sec{subsec:aprimereco} only makes use of kinematical information.  For large flight distance, the vector formed from the $pp$ collision to the \A decay vertex provides another measurement of the direction of flight.  Thus, a more sophisticated fitting procedure would not only constrain energy and momentum, but also enforce a consistent decay topology~\cite{Hulsbergen:2005pu}, likely improving the resolution.  In addition, during Run~1 LHCb managed to recover many of the bremsstrahlung photons emitted by electrons by reconstructing them in the calorimeter system~\cite{Aaij:1981106}.  These photons were then added back to the electron momenta.  Since the calorimeter occupancy will be higher in Run~3, we chose to assume that this recovery procedure would not be possible in our analysis.  If bremsstrahlung recovery is possible, though, it would improve the \mee resolution by $\mathcal{O}(10\%)$.  
We note that if a major upgrade of the calorimeter system\,\cite{Sheldon} is installed for Run~4 that utilizes large-area picosecond photodetectors\,\cite{LAPPDs}, this would significantly improve the \mee resolution.  Such an upgrade would also permit using many more $D^0$ decays.  

Installing additional tracking stations onto the face of the dipole magnet would provide LONG-track-level momentum resolution for UP tracks\,\cite{Sheldon}.  The hit resolution required is $\mathcal{O}({\rm mm})$, which means that such a tracking station would not be too costly.  Since many \A decays are reconstructed with at least one UP track, installing such a system could greatly improve the \mee resolution.  Furthermore, many $D^0$ decays produce UP tracks whose resolution would also be improved by installing a tracking system onto the face of the magnet.  

For simplicity, our resonant search imposed a conversion veto, but since the conversion background is smooth, one could still perform a resonant search without any $\ell_{\rm T}$ requirement at the expense of larger backgrounds.  This is particularly relevant for small $m_\A$ where the $D_{\rm T}$ buffer requirement is most inefficient.

Finally, one could consistently combine both displaced searches and the resonant search following a strategy similar to \Ref{Williams:2015xfa}.  This is particularly relevant if the \A can also decay into dark matter, since then the relationship between $\Gamma_{\A}$, $m_{\A}$, and $\epsilon^2$ given in \Eq{eq:gamma_a} will no longer hold and \A will have a shorter lifetime.  To avoid introducing model dependence into the \A search, one can define exclusive prompt and displaced regions experimentally, and then optimally combine the information from both regions regardless of the \A lifetime \cite{Williams:2015xfa}.  A combined strategy might also help close the remaining gap between the resonant and displaced searches at large $m_{\A}$.

\section{Comparison to Other Experiments}
\label{sec:other}

\begin{figure}[t]
\includegraphics[width=\columnwidth]{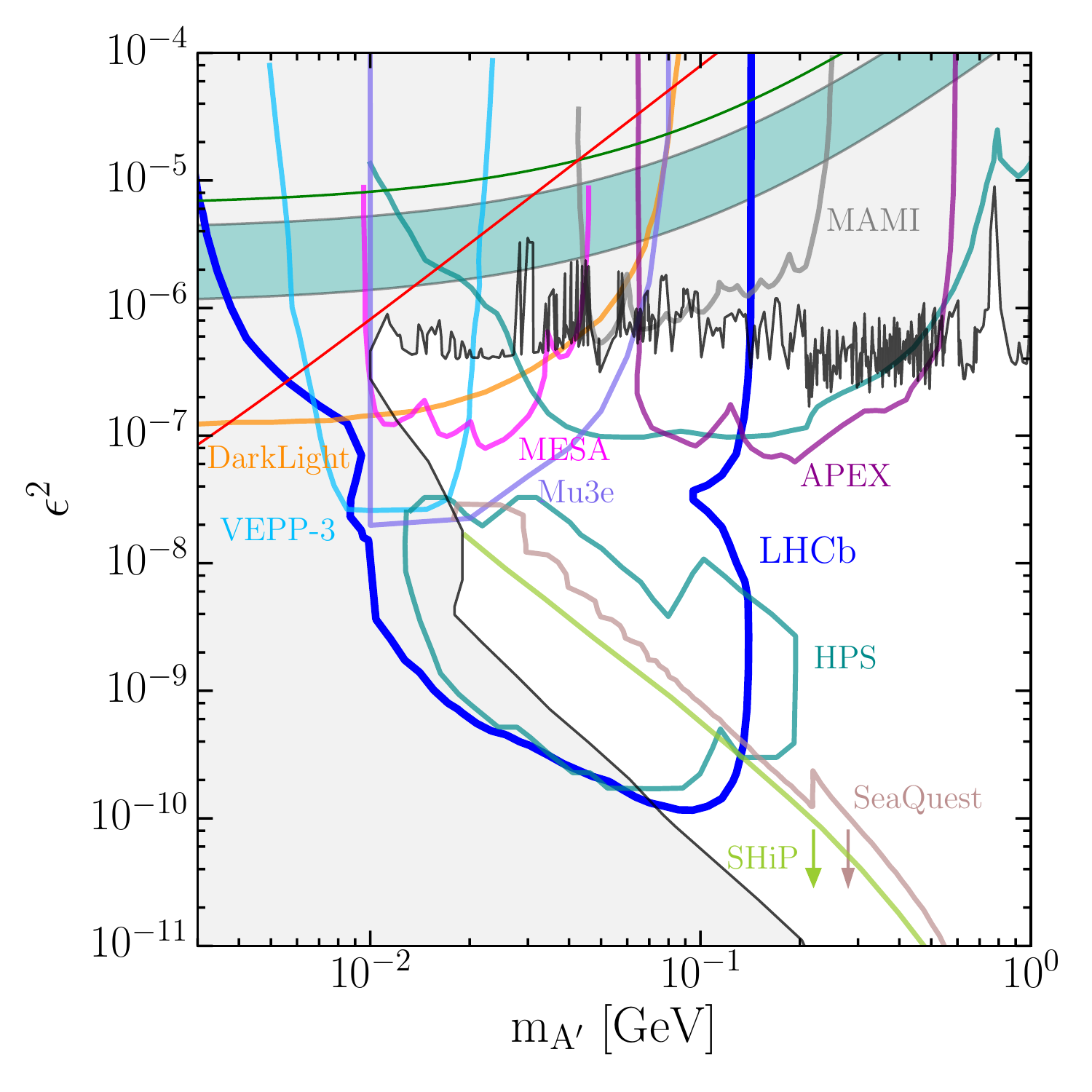}
\caption{Comparing the LHCb reach to other proposed dark photon experiments.}
\label{fig:lhcbboundswithoverlay}
\end{figure}

There is a rich planned program of dark boson searches \cite{Essig:2013lka}, and various experimental proposals are sensitive to the same dark photon mass and coupling range targeted by this \sigdecay search.   Their anticipated reaches are summarized in \Fig{fig:lhcbboundswithoverlay}.  Experiments like APEX \cite{Essig:2010xa, Abrahamyan:2011gv}, MESA/MAMI \cite{Beranek:2013yqa}, DarkLight \cite{Freytsis:2009bh,Balewski:2013oza}, VEPP-3 \cite{Wojtsekhowski:2012zq}, and Mu3e \cite{Echenard:2014lma} are high luminosity experiments that use a resonant search strategy.\footnote{Strictly speaking, VEPP-3 uses a missing mass strategy.}    Experiments like SHiP \cite{Alekhin:2015byh} and SeaQuest \cite{Gardner:2015wea} are beam-dump experiments that use a displaced strategy.  See \App{sec:belle} for discussion of a possible \sigdecay search at Belle-II.

The closest comparison to the LHCb \sigdecay search is the HPS experiment \cite{Moreno:2013mja}.  Because HPS has a dedicated tracking and vertexing detector, it is able to search for both resonant and displaced $\A$ signals, corresponding to the upper and lower HPS regions in \Fig{fig:lhcbboundswithoverlay}.  In terms of vertex performance and mass resolution, HPS is quite comparable to LHCb.\footnote{One minor difference is that HPS uses strips for its tracking while LHCb will use pixels in Run~3.  This means that HPS has worse hit resolution in the bending plane, so some of the topological requirements in \App{app:vertex} would not be helpful for HPS.}   It is therefore reasonable to ask why the resonant and displaced search regions overlap for LHCb but not for HPS.

There are three main advantages of LHCb over HPS, at least for $m_\A < 100$\:MeV.
\begin{itemize}
\item \emph{Parasitic running}.  For a fixed mass resolution, the resonant search is limited only by the available statistics.  The \sigdecay search does not require any modifications to the standard LHC running environment, so it immediately benefits from the high data-taking rate (and long run times) already needed by other LHCb measurements.  By contrast, HPS is a dedicated experiment with an anticipated runtime of only 3 weeks to cover the 30-100\:MeV mass range.
\item \emph{Access to smaller opening angles}.  As a fixed-target experiment, HPS produces $\A$ bosons in the very forward direction, effectively within the envelope of the beam pipe hole.  This means that HPS has no sensitivity in the ``dead zone'' where the \A decay opening angle is less than 30 mrad.  As a colliding beam experiment, LHC produces $\A$ bosons with a sizable transverse momentum kick, such that the $\A$ trajectory itself goes into the LHCb VELO.  This allows LHCb to reconstruct much smaller opening angles down to around a few mrad, which helps the reach at low $\A$ masses.
\item \emph{Larger Lorentz boosts}.  The reach in the displaced search benefits from large $\gamma$ factors (up to a point, see below).  The median $\A$ boost at the LHC is roughly three times larger than the maximum $\A$ boost at HPS.  Moreover, there is a tail of \A events at the LHC which extends to much higher boost factors, which can be exploited when combined with the high event rate.
\end{itemize}
Note that these last two bullet points are in direct conflict at a fixed-target experiment, since going to larger Lorentz boosts by using a higher beam energy means that the signal has smaller \A opening angles.  At a colliding beam experiment, the opening angle resolution is limited only by the hit resolution, so one can in principle exploit larger Lorentz boosts up until the point that the displaced $\A$ signal escapes the detector.   From this we conclude that HPS is probably close to optimal for a fixed-target dark photon search in this mass range.

\section{Summary}
\label{sec:conclude}

In this paper, we showed that in Run~3, LHCb can explore the entire dark photon parameter space between the prompt-\A and beam-dump limits for $m_{\A} \lesssim 100\mev$ using the decay \sigdecay.  This reach is possible due to the large $D^{*0}$ production rate and sizable \A Lorentz boost factor at the LHC, combined with the excellent vertex/mass resolution and planned triggerless-readout system of LHCb.  The displaced and resonant strategies give complementary coverage of the \A parameter space.  Even if the displaced vertex signature is absent due to a modified \A lifetime, there is still substantial coverage from the resonant search owing to the excellent \mee resolution.

Given the impressive reach below $\Delta m_D = 142\mev$, one might wonder if a similar search could be performed at LHCb for dark photons with larger masses.  The \sigdecay search relies on the $D^{*0}$ mass constraint to suppress backgrounds and to improve the \epem invariant mass resolution.  
Without these handles, the dark photon search becomes considerably more difficult, but there are a number of potential strategies to explore.

Above the dimuon threshold, a generic search could be performed for displaced $\A \to \mu^+\mu^-$ vertices that are inconsistent with originating from interactions with the detector material.   The dimuon invariant mass resolution is sufficient without applying a mass constraint, and data samples of displaced dimuon vertices are nearly background free at LHCb (see, e.g., \Ref{Aaij:2015tna}).  A similar search was performed by LHCb using Run~1 data that looked for the very rare decay $K_S^0\!\to\mu^+\mu^-$\,\cite{Aaij:1478934}.   A key challenge for these generic searches, though, is converting a cross section bound into a bound on $\epsilon^2$, since $pp$ collisions provide many potential sources of dark photons with uncertain production rates.

Other radiative charm decays may also offer viable \A search channels at LHCb.  For example, the yield of $D^0\!\to \kstarbar\gamma$ decays, where both the kaon and pion from the $\kstarbar\!\to K\pi$ decay are reconstructed as LONG tracks, will be $\mathcal{O}(10^{10})$ in Run~3.  A search could then be performed for $D^0\!\to \kstarbar\A$, with $\A\!\to\epem$, $\mu^+\mu^-$, and $\pi^+\pi^-$.   This search is sensitive to $m_{\A} \lesssim 1\gev$ and the $D^0$ mass constraint would ensure good \A mass resolution.
The \A lifetime is shorter at larger $m_{\A}$, however, so a displaced search would not be possible over most of the $m_{\A}$ range. Nevertheless, such a search is worth investigating.  Similarly, the decay $D_{s1}(2460)^+ \to D_s^+ \A$ may provide a viable search for $m_\A \lesssim 500\,\MeV$.

The common theme for all of the above search strategies is that they benefit greatly from triggerless readout with real-time data calibration.  These and other searches provide strong physics motivation for these upgrades to LHCb in Run~3.  We encourage more effort in exploring how best to exploit these advances in the hunt for dark photons and other new particles.

\section*{Acknowledgements}

We are indebted to Natalia Toro for extensive discussions on prompt backgrounds to the pre-module search.
We thank Rouven Essig, Marat Freytsis, Tim Gershon, Yoni Kahn, Patrick Koppenburg, Gordan Krnjaic, Zoltan Ligeti, Jia Liu, Iain Stewart, and Vincenzo Vagnoni for providing helpful advice and feedback.  
We thank Yasmine Amhis, Thomas Bird, Chris Jones, Marie-H\'{e}l\`{e}ne Schune, Paul Seyfert, Edwige Tournefier, and Daniel Vieira for help determining the Run~3 LHCb performance.  
J.T. and W.X. are supported by the U.S. Department of Energy (DOE) under cooperative research agreement DE-SC-00012567.  J.T. is also supported by the DOE Early Career research program DE-SC-0006389, and by a Sloan Research Fellowship from the Alfred P. Sloan Foundation.  J.T. thanks the Galileo Galilei Institute for Theoretical Physics for the hospitality and the INFN for partial support during the completion of this work.  M.W. and P.I. are supported by the U.S.\ National Science Foundation grant PHY-1306550.

\appendix

\section{Estimating \texorpdfstring{LHC\MakeLowercase{b}}{LHCb} Tracking Performance}
\label{app:res}

LHCb does not provide public fast simulation software.  Therefore in this appendix, we estimate the various performance numbers needed for our study from public documents.

We start with deriving the resolutions for LONG and UP tracks, as defined in \Sec{sec:tracks}.  The resolution on $\sigma_p/p$ for LONG tracks with $p > 10\gev$ as a function of $p$ in the current LHCb detector is provided in Fig.\:17 of \Ref{Aaij:2014jba}.  The LONG tracks used in this analysis typically have momenta in the region where
\be 
\sigma_p^{\rm LONG}/p \approx 0.5\%
\ee 
(n.b.\ $\sigma_p/p$ has no strong dependence on $p$ for $p<10\gev$).
The LONG track resolution is expected to improve by about 20\% in Run 3, but to be conservative we assume the performance will be as measured in the current detector.  
 The resolution for UP tracks with momenta typical of electrons coming from \A decays is 
\be
\sigma_p^{\rm UP}/p \approx 12\%,
\ee
as can be seen in Fig.\:2.5 in \Ref{LHCb-TDR-015}. 

The resolution on the track direction is not explicitly published in any LHCb document to our knowledge, but we can extract it from published mass resolutions.  LHCb measures $\sigma_p/p$ for tracks using various two-body decays, e.g.\ $\jpsi\!\to\mu^+\mu^-$.  For the case where $p_{\mu^+} \approx p_{\mu^-}$, the resolution on the track momentum is related to the resolution on the \jpsi mass $m$ and decay opening angle $\alpha$ by 
\be
\left(\frac{\sigma_p}{p}\right)^2 \approx 2\left(\frac{\sigma_m}{m}\right)^2 - 2\left(\frac{p\sigma_{\alpha}}{m}\right)^2,
\ee
which is the same as Eq.~(1) of \Ref{Aaij:2014jba} but without the additional factor of $\alpha$ in the denominator of the rightmost term (that is a typo in \Ref{Aaij:2014jba}).
Taking $\sigma_m = 14\mev$ from Tab.~2, along with $\sigma_p/p = 0.5\%$ and $p \approx 15\gev$ from Fig.\:17 of \Ref{Aaij:2014jba}, we obtain $\sigma_{\alpha} \approx 0.3$\:mrad, which gives a track polar-angle resolution of $\sigma_{\theta} \approx 0.2$\,mrad in the $\theta_{\jpsi}\to 0$ limit.  

The value $\sigma_{\theta} = 0.2$\,mrad should be considered valid only for tracks with large momenta.  For low-momentum tracks, multiple scattering dominates the resolution.  
We simulate a toy model of the Run~3 VELO taking the radiation length of the foil and detector modules from \Ref{LHCb-TDR-013}.  
The resulting resolution is 
\be
\label{eg:thetaresolution}
\sigma_{\theta} \approx \left( 0.2+ \frac{1.7\gev}{p} \right) \,{\rm mrad}, \quad  \sigma_{\phi} \approx \sigma_{\theta}\cot{\theta}.
\ee
When the \A is constrained to originate from the $pp$ collision, as assumed in deriving \Fig{fig:mee}, $\sigma_{\phi} \approx \left(\sqrt{\sigma_{\rm hit}^2 + \sigma^2_{pp,{\rm T}}}\,\,\right)\!/6\,{\rm mm} \approx 3$\,mrad (the hit and transverse $pp$ location resolution divided by the radial distance to the first hit).  

We now turn to the resolutions for F-type and P-type $D^0$ decays.  For two-body F-type $D^0$ decays, both tracks are LONG and so one would naively expect $\sigma_p/p \approx 0.5\% \times \sqrt{2}$.  However, applying the constraint of the known $D^0$ mass improves this by about a factor of two in our toy simulation.  In the $\theta_D \to 0$ limit, 
\be
\sigma_{\theta}^{D^0_F} \approx \frac{\sigma_{\alpha}}{2} \approx 0.2\,{\rm mrad}, 
\ee
with 
\be 
\sigma_{\phi}^{D^0_F} \approx \sigma_{\theta}\cot{\theta} \approx 2\,{\rm mrad}.
\ee  
For four-body F-type decays, the resolution is similar to the two-body case after applying the $D^0$ mass constraint, even if one or two of the tracks are of the UP type.  For simplicity, the largest $\sigma_p/p$ value obtained (when two tracks are UP)
\be
\sigma_p^{D^0_F}/p \approx 0.5\%
\ee
 is used for all F-type decays.  

For P-type $D^0$ decays, the resolution depends strongly on the $D^0$ flight distance $\ell$.  Therefore, improved resolution comes at the expense of signal efficiency.  
LHCb achieves a proper lifetime resolution of $\sigma_t \approx 50$\,fs in $D^0$ decays~\cite{PhysRevLett.110.101802}.  
Using the relationship
\be
\label{eq:sigl}
\sigma_t^2 = \left(\frac{m}{p}\right)^2\sigma_{\ell}^2 + \left(m \ell \right)^2 \left(\frac{\sigma_p}{p}\right)^2,
\ee
we can obtain a per-event estimate of the resolution on the flight distance. 
The vertex resolution in the plane transverse to the beam line ($x-y$) is much better than along the beam direction ($z$), e.g.\ in \Ref{Aaij:2014jba} $\sigma_z/\sigma_x \approx 5.5$ for the $pp$-collision vertex.  Therefore, the resolution on the transverse flight distance is $\sigma_{\ell_{\rm T}} \approx \sigma_{\ell} \sin{\theta}$, and $\sigma_{\theta} \approx (\sigma_{\ell_{\rm T}}/\ell\cos{\theta})(5.5/\sqrt{2}) \approx \tan{\theta}\sigma_{\ell}/4\ell$, where $\sigma_{\ell}$ is obtained from \Eq{eq:sigl} for each $D^0$ decay. Similarly, $\sigma_{\phi} \approx \sigma_{\ell}/4\ell$.  Both expressions have the expected scaling with flight distance.  Finally, applying $D^0$ selection criteria that are 50\% efficient gives 
\be
\sigma_{\theta}^{D^0_P} \approx 2\,{\rm mrad}, \quad  \sigma_{\phi}^{D^0_{\rm P}} \approx 30\,{\rm mrad}.  
\ee
In this toy study, we require $\sigma_{\theta} < 5$\,mrad and assume additional requirements, e.g.\ on particle identification, are applied that are 90\% efficient.  In a more sophisticated analysis where the full covariance matrix from the vertex fit is available, one could choose to cut on $\sigma_p/p$ as in \Ref{Aaij:2007377}.  

The $D^0$ momentum in P-type decays is solved for as described in \Sec{subsec:dreco}.  For small $\ell$ it is common that $p^{\perp}_{\rm mis}$ is unphysical, i.e.\ it is larger than the expected total missing momentum in the $D^0$ rest frame due to the resolution on the $D^0$ decay vertex.  In such cases, we vary the location of the $D^0$ vertex to find the position with the smallest vertex $\chi^2$ that provides a physical  $p^{\perp}_{\rm mis}$ value.  After requiring $\sigma_{\theta} < 5$\,mrad, we find 
\be
\sigma_p^{D^0_P}/p \approx 9\%.
\ee 
It is likely that a more sophisticated selection based on $\sigma_p/p$ would provide better resolution, though this has little impact on the search proposed in this paper.

\section{Estimating \texorpdfstring{LHC\MakeLowercase{b}}{LHCb} Vertex Performance}
\label{app:vertex}

The resolution on the IP and vertex location are key elements of the displaced searches.  The one-dimensional track IP resolution expected in Run~3 is well approximated by\,\cite{LHCb-TDR-013}
\be
\sigma_{\rm IP} = \left(11.0 + \frac{13.1\,{\rm GeV}}{p_{\rm T}} \right)\,\mu{\rm m}.
\ee
In the $\theta_{\A}\to0$ limit, this is the IP resolution both in and orthogonal to the \A decay plane.
We can use $\sigma_{\rm IP}$ to cross check our estimate of $\sigma_{\theta}$ for tracks.  Assuming that the location of the $pp$ collision is perfectly measured, then $\sigma_{\theta} \approx \sigma_{\rm IP}\sin{\theta}/6$\,mm, where 6\,mm is the mean radial distance of the inner-most track hit\,\cite{LHCb-TDR-013}. 
This gives $\sigma_{\theta} \approx 2.2\,{\rm GeV}/p$, which, for typical electron momenta in \sigdecay decays, agrees to within about 10\% with our estimate in \Eq{eg:thetaresolution} based on mass resolution and a toy VELO simulation.

\begin{figure*}[t]
\includegraphics[width=0.47\textwidth]{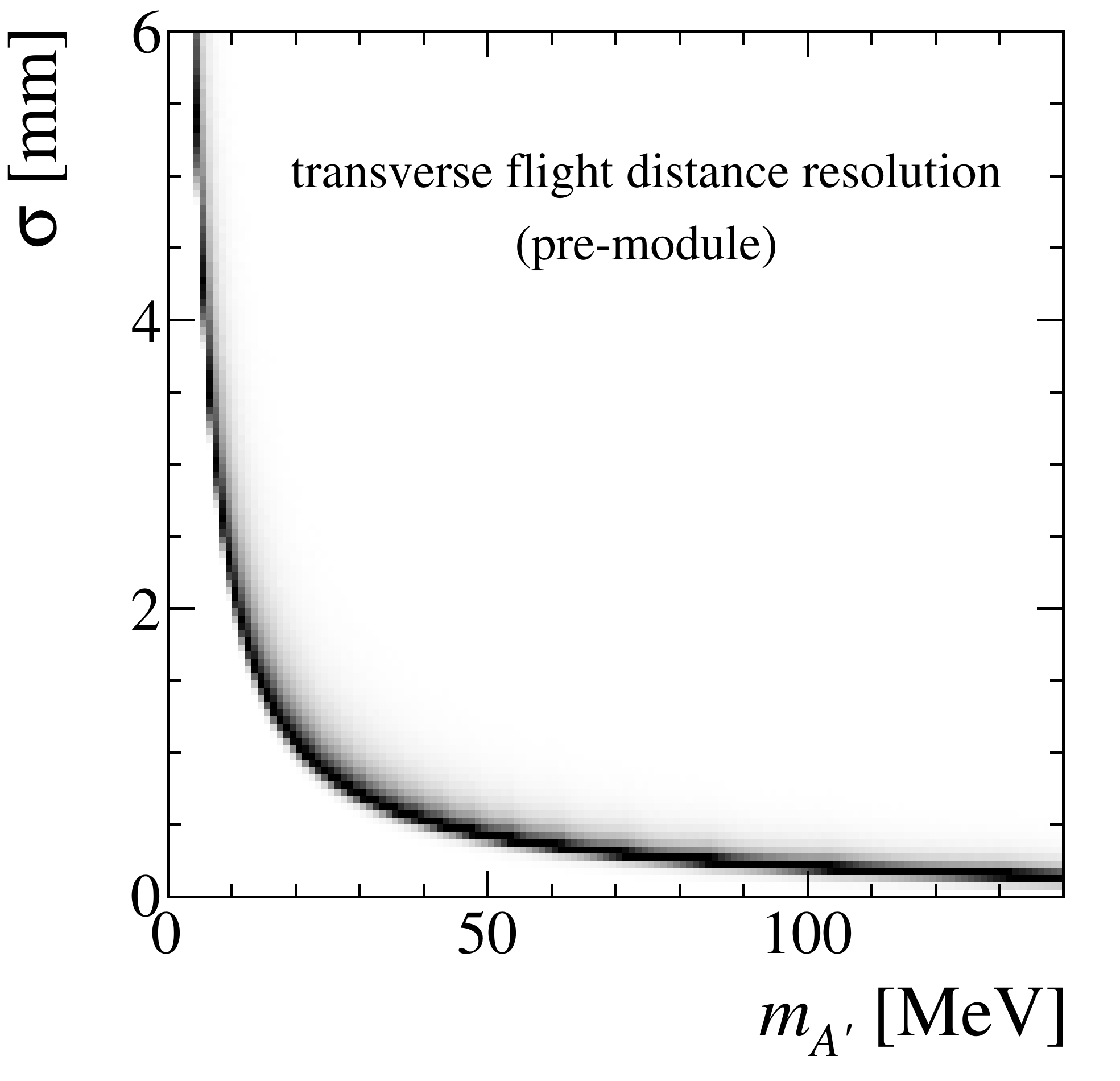}
\includegraphics[width=0.47\textwidth]{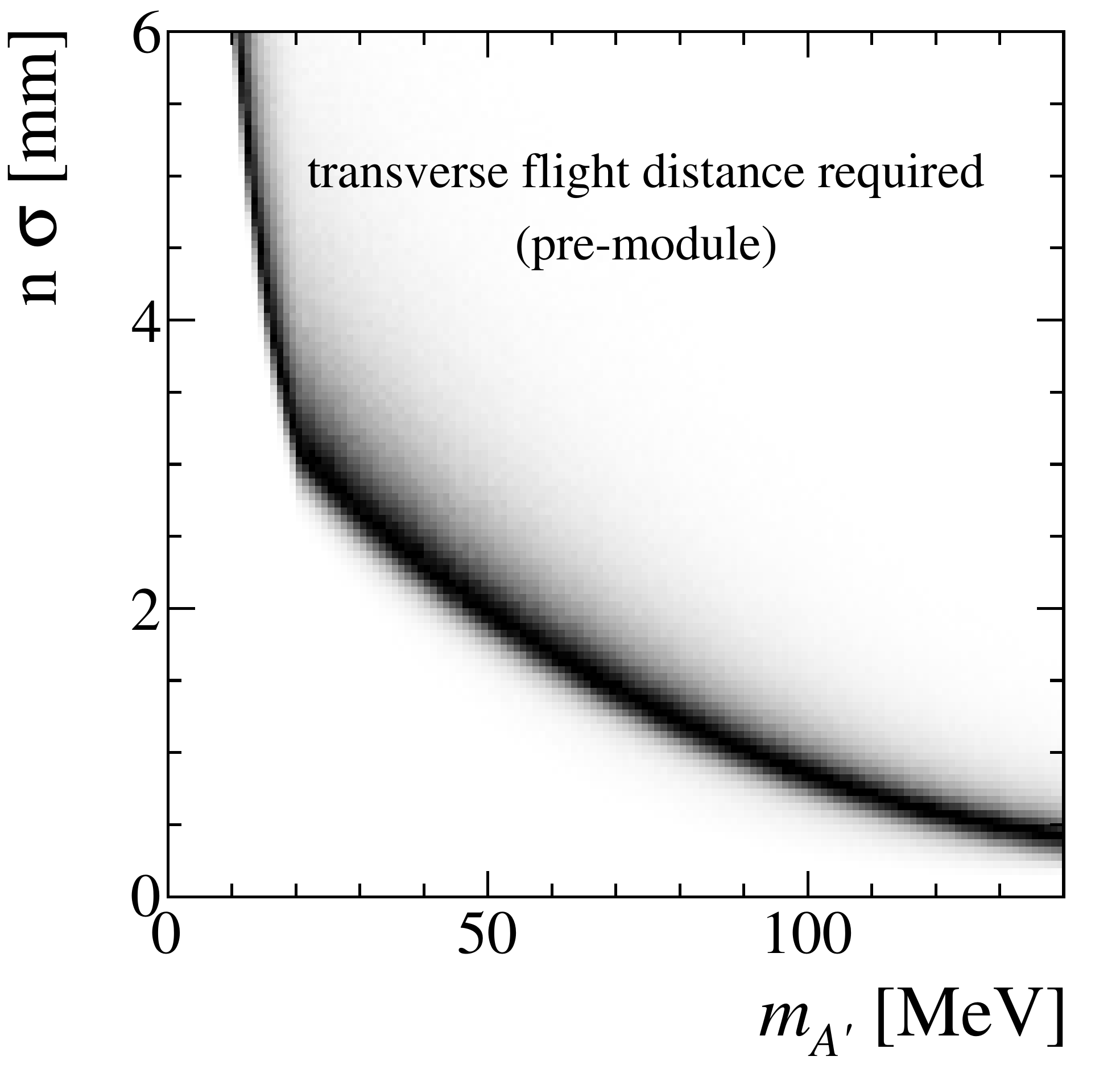}
\caption{Distribution of (left) resolution $\sigma_{\ell_{\rm T}}$ and (right) required displacement $n \, \sigma_{\ell_{\rm T}}$ versus $m_{\A}$ for pre-module \A decays.}
\label{fig:elltres}
\end{figure*}

The resolution on $\ell_{\rm T}$ in the pre-module search is well approximated by 
\begin{equation}
\sigma_{\ell_{\rm T}} \approx \frac{\sin{\theta}}{\alpha_{\epem}} \sqrt{\sigma_{e^+{\rm IP}}^2 + \sigma_{e^-{\rm IP}}^2},
\end{equation}
where $\alpha_{\epem}$ is the \A decay opening angle and the \A is constrained to originate from the $pp$ collision.  
The resolution on $\ell_{\rm T}$ as a function of $m_{\A}$ is shown in \Fig{fig:elltres}, as is the required displacement $n \, \sigma_{\ell_{\rm T}}$ to achieve $\approx 1$ background event in a given \A signal mass window.  The $m_{\A}$ dependence is driven by the dependence of $\alpha_{\epem}$ on $m_{\A}$.

In our pre-module displaced search, we require a consistent decay topology.  These requirements are as follows:
\begin{itemize}
\item the \A decay vertex is downstream of the $pp$ collision;
\item the distance of closest approach between the $e^+$ and $e^-$ tracks is consistent with zero;
\item the angle between $\vec{p}_{\A}$ and the vector formed from the $pp$ collision to the \A decay vertex is consistent with zero;
\item the IP out of the \A decay plane---defined by the $pp$ collision point and the first hits on the $e^+$ and $e^-$ tracks---for each electron is consistent with zero.
\end{itemize}
In each case, we define consistent with zero as having a $p$-value greater than 1\%.  Therefore, the efficiency on a true displaced \A decay is close to 100\%.  

An important point to keep in mind is that virtually all fake highly displaced \epem vertices  that satisfy these consistency requirements are reconstructed with $\alpha_{\epem}$ much larger than its true value.  Therefore, the reconstructed value of $m_{\epem}$ is larger than its true value in the absence of substantial bremsstrahlung, and the dominant background to the \A signal comes from (more copious) $e^+e^-$ pairs at smaller true invariant mass.

The opening angle in our search for $m_{\A} = 100\mev$ is on average about the same as for the HPS experiment.  
One can see from \Fig{fig:elltres} that the pre-module search requirement at 100\mev is $\ell_{\rm T} \gtrsim 1$\,mm or $\ell \gtrsim 25$\,mm.  
This value is similar to the flight distance requirement used by HPS to determine the reach of their displaced search, leading to a comparable high-side reach at that mass value (see \Fig{fig:lhcbboundswithoverlay}).

\begin{figure*}[t]
\includegraphics[width=0.47\textwidth]{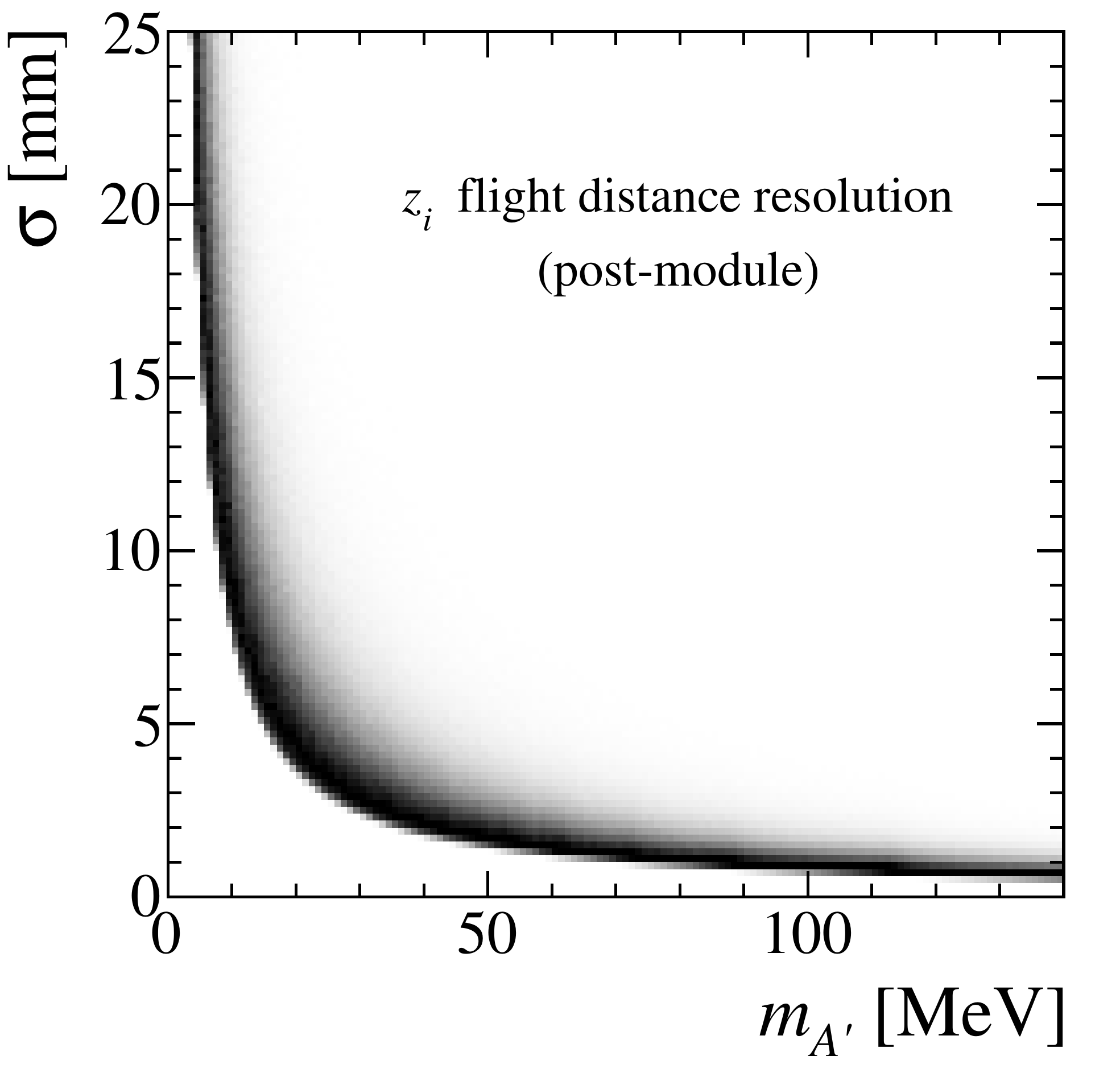}
\includegraphics[width=0.47\textwidth]{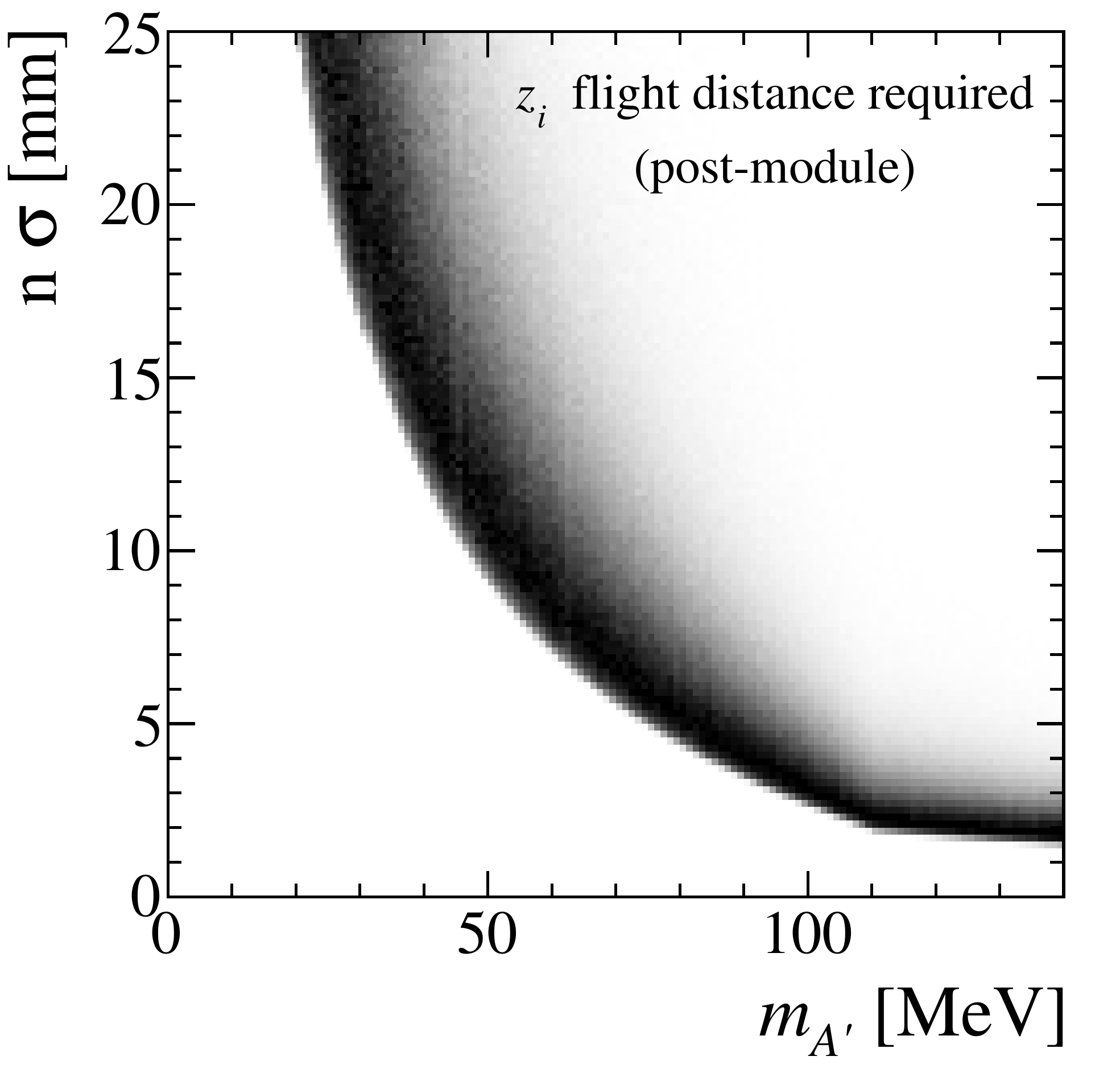}
\caption{Distribution of (left) resolution $\sigma_{z_i}$ and (right) required displacement $n \, \sigma_{z_i}$ versus $m_{\A}$ for post-module \A decays.}
\label{fig:zires}
\end{figure*}

For the post-module search, we define IP$_i$ with respect to the point where the \A trajectory intersects the $i^{\rm th}$ VELO module, which is the one directly upstream of the \A decay vertex location.  
The resolution on IP$_i$ is given by
\begin{equation}
\sigma_{{\rm IP}_i} \approx \sqrt{(\Delta Z_i \, \sigma_{\theta})^2 + (12\,\mu{\rm m})^2},
\end{equation}
where $\Delta Z_i$ is the separation of VELO modules $i$ and $i+1$, and 12\,$\mu$m is the hit resolution.  
For simplicity, we take $\Delta Z_i = 25$\,mm for all modules. In reality, about 40\% of the modules traversed by an \A candidate have a larger $\Delta Z_i$.  Such cases have slightly better resolution since the relative effect of the hit resolution is reduced.  
The resolution on the $z$ position of the \A decay vertex in the post-module search is
\begin{equation}
\sigma_{z_i} \approx \frac{1}{\alpha_{\epem}} \sqrt{\sigma_{e^+{\rm IP}_i}^2 + \sigma_{e^-{\rm IP}_i}^2},
\end{equation}
where again the \A is constrained to originate from the $pp$ collision.  
The resolution on $z_i$ is shown in \Fig{fig:zires}, as is the required displacement $n \, \sigma_{z_i}$.
The consistency requirements imposed for the post-module displaced search are the same as those used in the pre-module search, with the following exceptions: the \A decay vertex is required to be downstream of the $i^{\rm th}$ VELO module, and IP$_i$ is used in place of IP for the out-of-decay-plane requirement.

\section{\texorpdfstring{B\MakeLowercase{elle}}{Belle}-II Reach}
\label{sec:belle}

The search strategy in this paper can be applied to any experiment with a large $D^{*0}$ production rate.  As a key example, the Belle-II experiment~\cite{BelleII} will collect a data sample corresponding to about 50\:ab$^{-1}$ starting in 2017.  Indeed, Belle-II plans to collect about 10\:ab$^{-1}$ prior to the start of Run~3 at the LHC, so they may be able to set initial bounds on \sigdecay (or make a discovery) prior to LHCb.

Again neglecting secondary $D^{*0}$ production from $b$ hadron decays, the total number of prompt \smdecay decays produced at Belle-II will be approximately
\be
\hspace{-0.2in}N(\smdecay) \approx 510\,{\rm pb}\times 0.381 \times 50\,{\rm ab}^{-1} \! \approx\! 10^{10},
\ee
where the first number is the inclusive prompt  $e^+e^- \to D^{*0}$ cross section and the second number is the $\smdecay$ branching ratio.   This is about 500 times less than at LHCb in Run~3, though Belle-II will likely be able to make use of a larger fraction of $D^0$ decays than LHCb.  Rescaling the limits derived in this paper, we anticipate Belle-II will not be able to probe unexplored parameter space using the resonant $\A$ strategy. 

The main challenge for a Belle-II displaced search is that the electrons produced in \sigdecay decays will have very low momenta.  Using \pythia, we estimate that the median value of the lower-momentum electron is only 60\mev.  This is likely to reduce the tracking efficiency, the electron-momentum resolution, and the $e^+e^-$ vertex resolution.  The Lorentz $\gamma$ factors at Belle-II will also be much lower resulting in shorter flight distances.  We estimate that Belle-II could have sensitivity to displaced \A decays for small $m_{\A}$.  Given that such sensitivity may occur prior to Run~3 at LHCb or elsewhere, we encourage a detailed study by the Belle-II collaboration to assess the discovery potential.

\bibliography{ref}

\vspace{0.2in}

\end{document}